\documentclass[preprints,review,accept,moreauthors,pdftex,10pt,a4paper]{mdpi} 

\firstpage{1} 
\pubvolume{xx}
\issuenum{1}
\articlenumber{1}
\pubyear{2018}
\copyrightyear{2018}
\history{Received: ; Accepted: ; Published: }

\pdfoutput=1

\usepackage{amsfonts}
\usepackage{rotating}
\usepackage{pdflscape}
\usepackage{afterpage}
\usepackage{longtable}

\Title{Graphene-based light sensing: fabrication, characterisation, physical properties and performance.}

\Author{Adolfo De Sanctis$^{1,\ddagger,}$\orcidA{}, Jake D. Mehew$^{1,\ddagger,}$\orcidD{}, Monica F. Craciun$^{1}$\orcidB{} and Saverio Russo$^{1,}$*\orcidC{}}

\AuthorNames{Adolfo De Sanctis, Jake D. Mehew, Monica F. Craciun and Saverio Russo}

\address[1]{%
$^{1}$ \quad Centre for Graphene Science, College of Engineering, Mathematics and Physical Sciences, University of Exeter, Exeter EX4 4QL, United Kingdom;\\
}

\corres{Correspondence: s.russo@exeter.ac.uk;}

\secondnote{These authors contributed equally to this work.}

\abstract{Graphene and graphene-based materials exhibit exceptional optical and electrical properties with great promise for novel applications in light detection. However, several challenges prevent the full exploitation of these properties in commercial devices. Such challenges include the limited linear dynamic range (LDR) of graphene-based photodetectors, the lack of efficient generation and extraction of photoexcited charges, the smearing of photoactive junctions due to hot-carriers effects, large-scale fabrication and ultimately the environmental stability of the constituent materials. In order to overcome the aforementioned limits, different approaches to tune the properties of graphene have been explored. A new class of graphene-based devices has emerged where chemical functionalisation, hybridisation with light-sensitising materials and the formation of heterostructures with other 2D materials have led to improved performance, stability or versatility. For example, intercalation of graphene with FeCl$_3$ is highly stable in ambient conditions and can be used to define photo-active junctions characterized by an unprecedented LDR while graphene oxide (GO) is a very scalable and versatile material which supports the photodetection from UV to THz frequencies. Nanoparticles and quantum dots have been used to enhance the absorption of pristine graphene and to enable high gain thanks to the photogating effect. In the same way, hybrid detectors made from stacked sequences of graphene and layered transition-metal dichalcogenides enabled a class of detectors with high gain and responsivity. In this work we will review the performance and advances in functionalised graphene and hybrid photodetectors, with particular focus on the physical mechanisms governing the photoresponse in these materials, their performance and possible future paths of investigation.}


\begin{document}

\section{Introduction}
The discovery of graphene \cite{Geim2007} and more broadly of atomically thin materials has triggered a wealth of research in optoelectronics \cite{Koppens2014}, plasmonics \cite{Fei2012}, telecommunications \cite{Mueller2010}, solar energy harvesting \cite{Pospischil2014} and sensing \cite{Kim2014}. Several novel applications in these sectors exploit the unique combination of broadband absorption, ultrahigh ambipolar mobility and field effect tunability inherent to single layer graphene \cite{Novoselov2004} and its compatibility with unconventional substrates such as recent developments in textile electronics \cite{Neves2015,Neves2017}. However, the lack of a bandgap and the intrinsic low light absorption of this single layer of carbon atoms poses some challenges for its use in practical applications \cite{Craciun2011}. Chemical functionalisation \cite{Craciun2013} has been proposed as a route to engineer an energy gap in the energy dispersion of graphene. At the same time, the ability to combine different two-dimensional (2D) materials into van der Waals (vdW) heterostructures \cite{Geim2013} has transformed this field of research owing to the possibility to create clean interfaces among system with very diverse physical properties such as semiconductors, insulators, superconductors, magnetic materials and ferroelectrics to list a few \cite{Novoselov2016}.

Photodetectors are used in nearly every electronic device which  interfaces with the external world or with other devices. Several industrial sectors make use of light sensors such as telecommunications, food production, transport, defence and healthcare. Although the miniaturisation of electronic devices such as transistors has allowed higher computational speeds and smaller, portable devices, the miniaturisation of light sensors did not proceed at the same rate, as several physical factors limit the scaling of such devices. In particular, the realisation of ultra-thin and flexible photodetectors is particularly challenging with conventional semiconductor technologies, due to the brittle nature of the materials used and low absorption at nano-scale thickness. Graphene-based light sensors have shown some exceptional performances, spanning from high speed \cite{Mueller2010} to large linear dynamic range (LDR) \cite{DeSanctis2017}, and they offer transparency and flexibility for future applications in wearable electronics \cite{Neves2015}. In this work, we will review the progress in the fabrication and characterisation of photodetectors (PDs) using chemically functionalised forms of graphene and hybrid graphene heterostructures with nanoparticles (NPs), transition-metal dichalcogenides (TMDs) and organic crystals. Whilst functionalisation can be used to efficiently modify the charge carrier dynamics in graphene which in return can lead to enhanced photoresponse, the hybridisation with NPs, TMDs and organic semiconductors boosts the absorption of light, therefore increasing the efficiency of the PDs. After a description of the materials and fabrication techniques, we will focus our attention on the main physical mechanisms responsible for photodetection in these materials. We will then review the most relevant papers which demonstrate their performance, highlighting strong and weak points for each device, as well as their suitability in specific applications.
 
\section{Materials and fabrication of graphene-based photodetectors}
Graphene can be obtained via different methods. The first and most direct approach is micro-mechanical cleavage of bulk graphite \cite{Novoselov2004}. This method gives very high quality single- and multi-layer graphene flakes, with very high values of mobility and low defect densities. Exfoliated graphene is often the starting material for functionalisation or for the creation of hybrids and heterostructures. However, this method is not scalable and has a low throughput. More scalable methods to produce graphene have been developed with chemical vapour deposition (CVD) being the most promising technique to produce large-area graphene \cite{Bointon2015CVD}. Depending on the substrate, CVD graphene can be grown as a single-layer or as a multi-layer. In its multi-layer form, CVD graphene is well suited to intercalation. Another scalable technique is liquid-exfoliation of graphite to produce graphene dispersion in water or other solvents \cite{Hernandez2008}. This route is often used to produce multi-layer graphene depositions, for example using vacuum filtration. Other production techniques include epitaxial growth on silicon carbide \cite{Bointon2014} and reduction of graphene oxide (GO), a functionalised form of graphene.

\begin{figure}
	\centering
	\includegraphics[width= 150 mm]{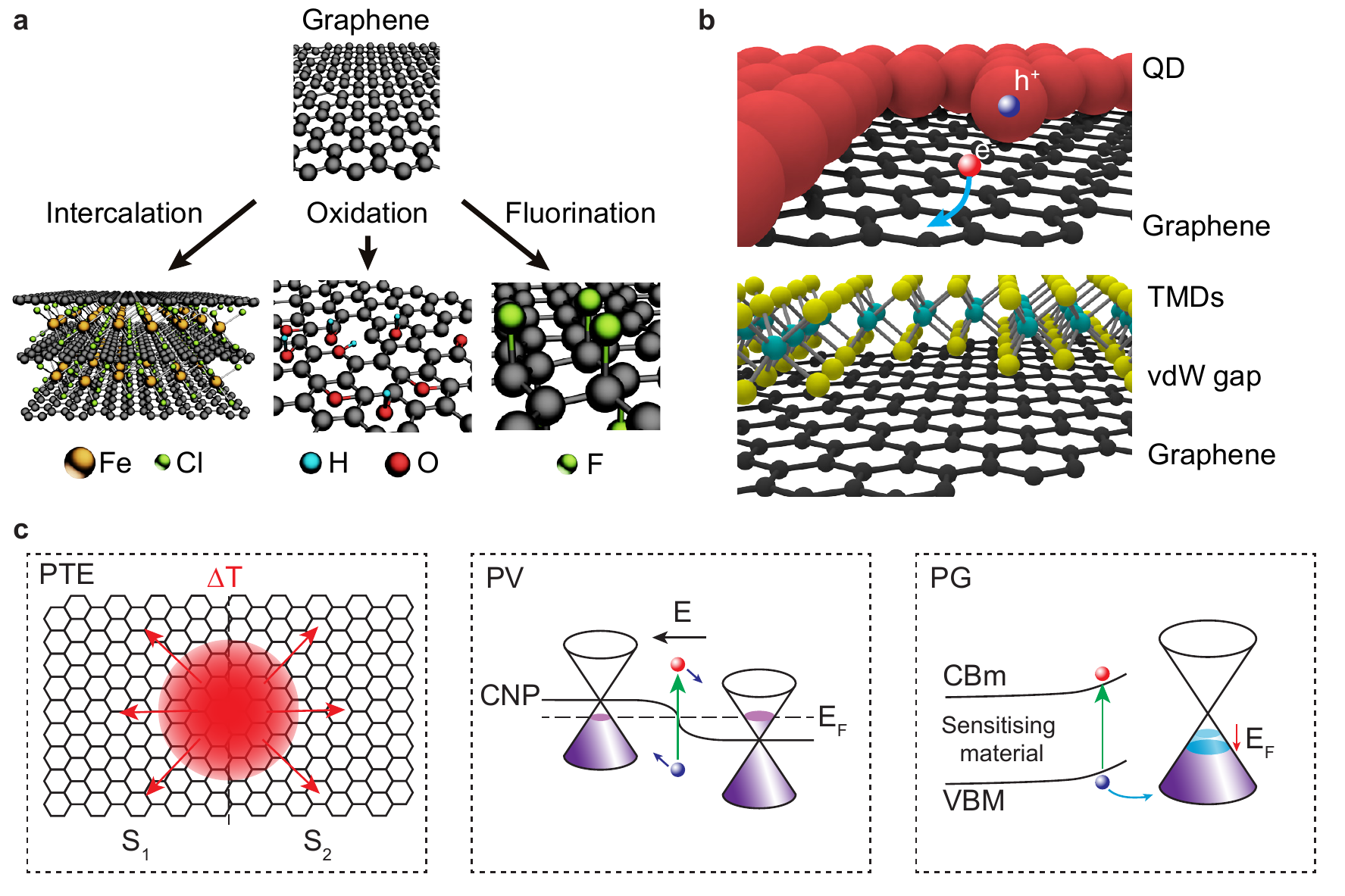}
	\caption{\textbf{Materials and photodetection mechanisms in graphene-based devices.} (\textbf{a}) Examples of functionalised graphene materials. (\textbf{b}) Examples of hybrid/heterostructure graphene materials: quantum dots (QD) and van der Waals (vdW) heterostructures made with graphene and semiconducting TMDs. (\textbf{c}) Three main mechanisms responsible for photo-activity in graphene-based devices: photothermoelectric effect (PTE), photovoltaic effect (PV) and photogating (PG). $S_1,\, S_2$ represent Seebeck coefficients, $T$ is the temperature, CNP is the charge neutrality point, $E_F$ it the Fermi level, $E$ is the electric field, VBM is the valence band maximum and CBm is the conduction band minimum.}
	\label{fig:Figure1}
\end{figure}

Chemical functionalisation refers to the use of chemical species to modify the properties of a material and in graphene it can take different forms \cite{Craciun2013}, which are illustrated in figure\,\ref{fig:Figure1}a. These are (1) the intercalation of chemical species between the layers of graphene, as with FeCl$_3$ \cite{Khrapach2012}, (2) the substitution of a carbon atom with an atom of a different specie or molecules, as with graphene oxide \cite{Mkhoyan2009}, or (3) the chemisorption of chemical species to saturate the $\pi$ bonds, as with fluorographene \cite{Robinson2010}. The functionalisation routes described can be realised using both solution- (e.g. sol-gel, hydrothermal, hydrolysis, solvent/ion exchange) and dry-process methods (e.g. plasma sputtering, annealing, thermal evaporation, CVD and PVD). Different techniques result in different size of the material, different defects and contaminants as well as different functionalization. For example, graphite oxide is prepared in solution, typically following Hummers' method, with GO isolated through liquid-phase exfoliation of the bulk material. Such a method is able to sustain a high production throughput as opposed to the oxgyen plasma treatment of CVD grown graphene, also known to produce GO \cite{Alexeev2017}. Each type of functionalisation produces materials with different electronic properties, spanning from metallic \cite{Khrapach2012,Bointon2015,Bointon2015a} to a wide-bandgap insulator and magnetic systems \cite{Bointon2014}. A functionalised graphene PD is generally made by contacting the active material with metal contacts, either defined by lithography or through the use of a shadow mask. Sometimes encapsulation is necessary due to the environmental instability of some of the constituent materials. This is usually achieved by depositing a polymer layer on top, such as Poly-methyl-methacrylate (PMMA), or by using an ionic liquid \cite{Ye2012,Ye2011,Ye13002} such as lithium perchlorate-PEO (LiClO$_4$-PEO) which also acts as a gating electrode \cite{Robin2016,Mehew2017}.

A multitude of techniques can be employed in order to form hybrids and heterostructures between graphene and other materials. These include coating techniques such as spin/spray/dip/cast/bar coating, printing techniques such as inkjet and contact printing and deposition methods such as thermal evaporation, CVD, electrochemical deposition, etc. However, the majority of these methods result in a certain degree of defects being introduced in the graphene material. In this review, we focus our attention onto non-destructive techniques which allow the realisation of heterostructures which preserve the high-mobility of graphene field-effect transistors (FETs) and enhance the light absorption of the device. For this reason we will consider the following two techniques: spin-coating (or equivalent deposition method) of nanoparticles or quantum dots (QDs) directly on the surface of graphene and stacking different 2D materials in a van der Waals (vdW) assembly \cite{Geim2013}, as shown in figure\,\ref{fig:Figure1}b. In both cases charges are extracted from the graphene layer by means of metal contacts. Encapsulation of vdW heterostructures is usually achieved using hexagonal Boron Nitride (hBN) \cite{Mayorov2011}, PMMA or sputtered oxides such as SiO$_2$ or AlO$_2$. Encapsulation in hBN allows the formation of one-dimensional contacts \cite{Wang614} (also known as side-contacts), which have been proved to give the lowest contact resistance whilst preserving the intrinsic properties of graphene resulting in record high charge carrier mobility, without any high temperature annealing steps common to high quality graphene devices.

In the context of materials for light-sensing applications, it is important to notice the difference between functionalisation, which refers to changes in the structure or nature of the host material, and hybridisation, which refers to property combination of two or more materials. The former results in a new material which is directly used as both light-absorber and charge-transport layer. In the latter, generally, one material acts as light-absorber, whilst the other, graphene in this case, acts as charge-transport layer.

\section{Light detection in graphene-based devices}
The main different mechanisms responsible for the photoresponse of functionalised and hybrid graphene PDs can be grouped in three categories: photothermoelectric (PTE), photovoltaic (PV) and photogating (PG) effects. These three mechanism are schematically shown in figure \ref{fig:Figure1}c. They all rely on the creation of a non-equilibrium distribution of photo-excited carriers and their diffusion or drift in a potential gradient. Although they can all be present in one device, one of them is dominant depending on the geometry and the microscopic carriers dynamic. Such dominant effect dictates the performance of the photodetector and its range of technological applicability. 

Other three mechanisms responsible for photodetection in graphene devices are the bolometric effect \cite{Richards1994}, the Dyakonov-Shur effect \cite{Dyakonov1996} and plasmon-assisted photocurrent generation \cite{Lundeberg2016}. These effects are particularly suited to detect mid-infrared (MIR) to THz radiation, however they will not be considered in this review as they are not dominant mechanisms in the functionalised and hybrid graphene PDs under consideration. Other methods to enhance the absorption of pristine graphene, such as coupling to plasmonic structures \cite{Polyushkin2013,Cai2015}, wave-guiding \cite{Wang2013} and micro-cavity resonators \cite{Engel2012} will not be considered since these techniques rely on engineering the substrate or the device rather than modifying the active material. For an overview on how nanophotonics structures are used in graphene photodetectors we suggest the review by Xia \textit{et al} \cite{Xia2014} and references therein.

\subsection{Characterisation and figures of merit}
The basic characterisation techniques rely on shining light onto the device whilst recording its electrical response. Light can impinge on the whole surface of the device, known as flood illumination, or it can be delivered with a focused laser onto a specific area to allow for a spatially-resolved photo-response, such as in scanning photocurrent mapping (SPCM) \cite{DeSanctis_G16}. Both techniques give insight on the physical nature of the observed photoresponse. With these techniques it is possible to extract the key quantities which define the performance of a photodetector, which are summarised in table \ref{tbl:SummaryPD}. 

\begin{table}
	\caption[Photodetector parameters]{\textbf{Summary of parameters used to characterise PDs.} Coupling Factors and wavelength dependence of all quantities are omitted for clarity of notation.}
	\label{tbl:SummaryPD}
	\begin{center}
		\begin{tabular}{lccc}
			\hline
			{\centering Quantity}  & 
			{\centering Symbol} & 
			{\centering Definition\textsuperscript{\emph{a}}}&
			{\centering Units}\\
			\hline
			External Quantum Efficiency	&	$\eta_e$, EQE	& $(I_{ph}/q)/\phi_{in}$ & \%\\
			Internal Quantum Efficiency	&	$\eta_i$, IQE	& $(I_{ph}/q)/\phi_{abs}$ & \%\\
			Operating Bandwidth & $\Delta f$ & - & $\mathrm{Hz}$ \\
			Gain & G & $\left(\mu \tau E\right)/L$ & -\\
			Responsivity	&	$\mathfrak{R}$	& $I_{ph}/P_{opt}$ & A/W (V/W)\\
			Noise Equivalent Power & NEP & $S_n/\mathfrak{R}$ & W$/\sqrt{\mathrm{Hz}}$ \\
			Specific Detectivity & $D^\star$ & $(A \Delta f)^{0.5}/\mathrm{NEP}$ & cm$\sqrt{\mathrm{Hz}}$/W\\
			Linear Dynamic Range & LDR & $10\times\log_{10}\left(P_{sat}/\mathrm{NEP}\right)$ & dB \\
			\hline
		\end{tabular}
	\end{center}
	
	\textsuperscript{\emph{a}} $I_{ph}=$ Measured photocurrent, $\phi_{in}=$ Incident photon flux, $\phi_{abs}=$ Absorbed photon flux, $P_{opt}=\phi_{in}/S =$ Incident optical power density, $A=$ Device area, $S_n=$ Noise spectral density, $P_{sat}=$ Saturation power, $L=$ channel length, $\tau=$ lifetime of photoexcited carriers, $E=$ electric field;
\end{table}

The responsivity $\mathfrak{R} = I_{ph}/P_{opt}$ is defined as the ratio between the measured photocurrent (or photovoltage) $I_{ph} = I - I_{dark}$, where $I_{dark}$ is the dark current, and the incident optical power $P_{opt}$ and it is measured in units of A/W (V/W). Noise in photodetectors plays an important role in real-life applications, the main figure of merit for the characterisation of noise is the Noise Equivalent Power (NEP), defined as the incident power necessary to produce a signal-to-noise ratio of $1$ at $1\,\mathrm{Hz}$ bandwidth. It is given by the noise spectral density $S_n$ divided by the responsivity: $\mathrm{NEP} = S_n/\mathfrak{R}$ and measured in units of W$/\sqrt{\mathrm{Hz}}$. The bandwidth $\Delta f$ of a PD is defined as the frequency at which its output power drops by $1/2$, that is when the photocurrent drops by $\sim 70.7\%$ (known as $-3\,\mathrm{dB}$ bandwidth). These quantities are used to define the main figure of merit in PDs performance, the specific detectivity $D^* = \left(A \Delta f\right)^{0.5}/\mathrm{NEP}$, where $A$ is the area of the device. $D^*$ is measured in \textit{Jones} (cm$\sqrt{\mathrm{Hz}}$/W) \cite{Jones1960}. The linear dynamic range (LDR) determines the region of incident power within which the photodetector has a linear response \cite{DeSanctis2017}. It is defined as the logarithm of the ratio between the saturation power $P_{sat}$ (at which the response of the detector deviates from linearity) and the NEP: $\mathrm{LDR} = 10\times\log_{10}\left(P_{sat}/\mathrm{NEP}\right)$. A final figure of merit is the gain which depends on the mobility $\mu$, the photoexcited carriers lifetime $\tau$ and the applied electric field $E$: $\mathrm{G} = \left(\mu \tau E\right)/L$.

\begin{figure}
	\centering
	\includegraphics[width= 150 mm]{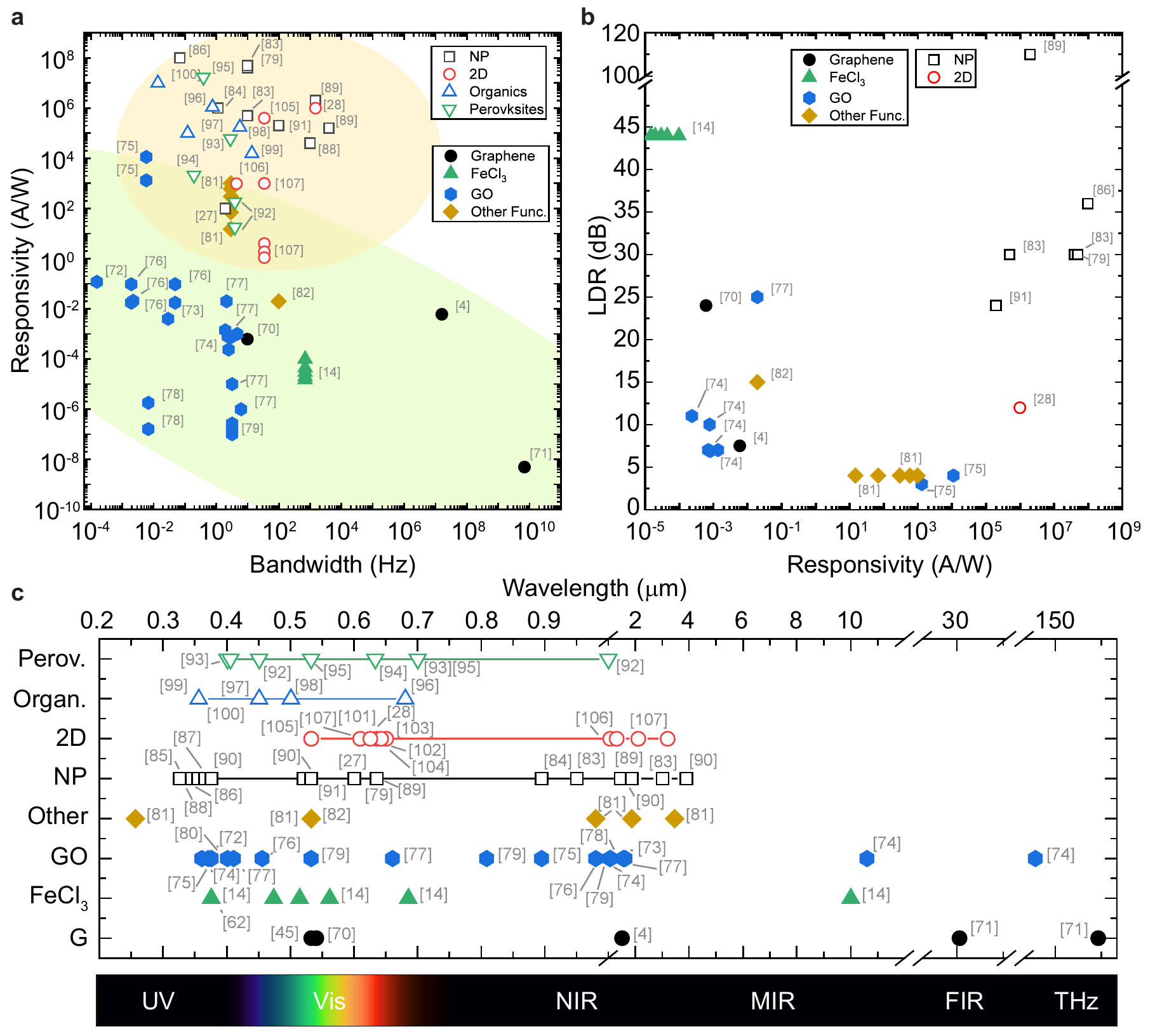}
	\caption{\textbf{Graphene-based photodetectors performance comparison.} (\textbf{a}) Responsivity \textit{vs} bandwidth and (\textbf{b}) LDR \textit{vs} responsivity in functionalised (filled symbols) and hybrid (open symbols) graphene PDs. Shaded areas encircle the majority of points belonging to each group. (\textbf{c}) Operational wavelength range for different graphene-based PDs, points correspond to experimentally tested wavelengths, lines represent full spectral scans. Detailed data in table \ref{tbl:SummaryPDsPerformance}, reference numbers in brackets. NP = nanoparticles, 2D = TMDs/heterostructures.}
	\label{fig:Figure2}
\end{figure}

Figure \ref{fig:Figure2} and table \ref{tbl:SummaryPDsPerformance} show a summary of the performance of the devices considered in this review. In general, in a plot of responsivity \textit{vs.} bandwidth, we can see a net separation between functionalised graphene and hybrid PDs, with the former having lower responsivity values than the latter (shaded areas in figure \ref{fig:Figure2}a). However, large LDR is observed in some functionalised graphene detectors, albeit with low values of responsivity. High LDR and high responsivity are both found in hybrid graphene PDs, thanks to the low NEP found in such devices. In terms of spectral response, figure \ref{fig:Figure2}c shows that both type of detectors are suited to a very wide range of incident photon energy. Of particular relevance is the ability of GO-based PDs to operate from UV to THz frequencies.

\subsection{Photothermoelectric effect}
 A junction between two materials with different Seebeck coefficients $S_1$ and $S_2$, in which the two sides are held at different temperatures, is subject to a voltage, known as thermoelectric voltage \cite{Seebeck1826}. In graphene, absorbed photons create a population of carriers with an increased temperature with respect to the surrounding. Such temperature gradient $\Delta T$, in the presence of a boundary to a material with a difference in Seebeck coefficient $\Delta S = S_1 - S_2$ (see figure \ref{fig:Figure1}c), causes a photovoltage to be generated:

\begin{equation}
	\label{eq:PTE}
	\Delta V = \Delta T \cdot \Delta S,
\end{equation}
where the signs is dictated by either gradients. This is known as photothermoelectric effect (PTE).  The Seebeck coefficient can be expressed using the Mott relation\cite{Mott1969,Hwang2009}:

\begin{equation}
	\label{eq:Seebeck}
	S = \frac{2\pi^2k_BT_{e}}{3qT_F},
\end{equation}
where $T_e$ is the electron temperature, $T_F =  E_F/k_B$ is the Fermi temperature, $q$ is the electron charge and $k_B$ is the Boltzmann constant. Equation \ref{eq:Seebeck} assumes that the mobility does not depend on the Fermi energy $E_F$, i.e. at relatively large doping. For $E_F \simeq 0.1 \,\mathrm{eV}$ the Seebeck coefficient has a value $S \sim 0.1\,\mathrm{mV/K}$ \cite{Tielrooij2015}. In general, the amount of energy taken by the hot carriers is given by $C_h T_h \propto P_\mathrm{opt}$, where $C_h$ is the heat capacity and $P_\mathrm{opt}$ is the incident optical power. Assuming the hot carriers thermalise at a temperature far above that of the lattice, their specific heat is $C_h\propto T_h^2$ and, combining equations \ref{eq:PTE} and \ref{eq:Seebeck}, the generated photocurrent is: $I_\mathrm{PTE}\propto T_h^2$. Therefore the proportionality between the generated photocurrent and the incident optical power is:

\begin{equation}
\label{eq:PTEProportional}
I_\mathrm{PTE} \propto \left(P_\mathrm{opt}\right)^{\frac{2}{3}}.
\end{equation}

The exponent in equation \ref{eq:PTEProportional} is commonly measured in graphene photodetectors, however a range of other exponents are possible due to PTE, depending on the dominant cooling mechanism. In particular an exponent of $1$ is possible if the electronic temperature is only marginally above the lattice temperature \cite{DeSanctis2017}.

It has been shown that the PTE is responsible for the light-sensing ability of pristine graphene and that the photo-active areas are confined to the junctions between two different materials such as the graphene/metal \cite{Mueller2010,Lemme2011,Echtermeyer2014}, monolayer/bilayer graphene \cite{Xu2010,Gabor2011,Kim2016} and graphene/functionalised graphene interfaces \cite{Withers2013}. Furthermore, the type of substrate and gate dielectric in graphene FETs has been shown to change the PTE properties of pristine graphene devices \cite{Xiangquan2016}. The hot-carrier dynamic in graphene has been extensively studied \cite{Bistritzer2009,Song2011,Song2012,Tielrooij2013} and the ability of graphene to generate large PTE voltages related to its unique band structure and high Fermi velocity ($v_F \simeq 10^6\,\mathrm{m/s}$) which limit the available states in the Fermi sphere for acoustic phonon scattering (responsible for cooling). Consequently, this limits the energy dissipation for photo-excited carriers, creating a population of electrons with a large effective temperature. Although PTE has been shown to dominate in graphene junctions devices, the presence of photovoltaic (PV) effects cannot be excluded, as it has been recently reported in single/multi-layer graphene junctions \cite{Zhang2018}.

\subsection{Photovoltaic effect}
The term "photovoltaic effect" has been commonly used by the solar cell research community to describe a broad variety of mechanisms by which the absorption of photons, generation of excitons, separation into free charge carriers and collection of free charge carriers at electrodes sequentially takes place. This definition is quite broad and does not necessarily require the presence of an electric field. However, within the research field of atomically thin materials the PV effect refers to the process of separation of photo-generated carriers by a built-in electric field \cite{Gabor648,Lemme2011,Freitag2012,Koppens2014}. These charges are subsequently extracted by a diffusion process in the short-circuit configuration or accelerated by an applied electric field to the electrodes. In this case, the resulting photocurrent is equal to \cite{WilsonOptoEl}: 

\begin{equation}
	\label{eq:PhotocurrentPhotoCond}
	I_{ph} = \frac{\eta_i P_{opt}}{\hbar \omega_0} \frac{W}{L} \tau_c q \left(\mu_e + \mu_h\right) V_{sd},
\end{equation}
where $\eta_i$ is the internal quantum efficiency (IQE), $\hbar \omega_0$ is the energy of the photon, $W$ and $L$ are the width and length of the device, $\tau_c$ is the carriers lifetime, $\mu_e$ and $\mu_h$ are the electron and hole mobilities and $V_{sd}$ is the applied bias. Equation \ref{eq:PhotocurrentPhotoCond} shows that the observed photocurrent has a linear dependence with respect to the incident optical power: $I_{ph} \propto P_{opt}$.

\subsection{Photogating and gain mechanism}
To increase the absorption of graphene-based photodetectors a semiconducting material is placed in close proximity to the channel of a graphene field-effect transistor (FET). Upon illumination a photoexcited charge carrier is transferred from the semiconductor to graphene. This changes the carrier density in the graphene FET which manifests itself in electrical measurements as a shift in the charge neutrality point ($V_{CNP}$) - effectively a photo-activated gate, hence the name photogating effect (PG). 

Such a system can be treated as a photoconductor with distinct light-absorbing and current-carrying regions. The photocurrent ($I_{ph}$) flowing in a device of area $A=WL$ and thickness $D$ is described by \cite{Sze_Semiconductors}:

\begin{equation}
	I_{ph} = (\sigma E)WD = (q\mu n E)WD,
	\label{eq:ipc}
\end{equation}
where $\sigma$ is the conductivity, $E$ the electric field across channel and $\mu$ the mobile carrier mobility. With the following definition for the number of photogenerated carriers ($n$);

\begin{equation}
	n = \frac{\eta(P_{opt}/h\nu)\tau}{WLD},
	\label{eq:dens}
\end{equation}
which includes the number of incident photons ($P_{opt}/h\nu$), quantum efficiency ($\eta$) and recombination rate ($1/\tau$). By using the earlier definition of responsivity ($\mathfrak{R}=I_{ph}/P_{opt}$) we arrive at:

\begin{equation}
	\mathfrak{R} = \left(\frac{q}{h\nu}\right)\eta\left(\frac{\mu\tau E}{L}\right) = \left(\frac{q}{h\nu}\right)\eta \mathrm{G}.
	\label{eq:resp}
\end{equation}

The responsivity of a typical hybrid graphene photodetector depends on three terms: the first is comprised of physical constants whilst the second and third terms relate to the quantum efficiency and gain of the system respectively, both of which need to be maximised.

Light will be absorbed by a semiconductor if the incident photons have energy greater than the band gap ($h\nu\ge E_g$). In this case electron-hole pairs are generated which form an exciton with an intrinsic efficiency ($\eta_{gen}$) that relates to the absorption coefficient of the material. To create free charges, the Coulomb force between electron and hole must be overcome. This can happen under the influence of large electric fields or due to thermal energy and this process has an associated efficiency term ($\eta_{diss}$). Charges are transferred between semiconductor and graphene in the presence of a potential barrier at the semiconductor-graphene interface or from a charge trapping mechanism in the semiconductor. In addition clean interfaces are required for efficient charge transfer ($\eta_{trans}$). Therefore the quantum efficiency can be split into three terms:

\begin{equation}
\eta = \eta_{gen}\eta_{diss}\eta_{trans}.
\label{eq:eta}
\end{equation}

Applying a bias voltage to the graphene channel allows the transferred charge to be extracted at the drain contact. To preserve electrical neutrality a charge must be simultaneously injected at the source. This process of charge recirculation can occur multiple times until the recombination of the trapped charge takes place. Such a process is described by the gain term in equation \ref{eq:resp}. To achieve the largest gain the ratio between the trapped carrier lifetime ($\tau$) and free carrier transit time ($t_{tr}=L/\mu E$) must be maximised: $\mathrm{G} = \tau/t_{tr}$.

Long-lived charge trapping is achieved by the spatial separation of photoexcited charges across the interface. $\tau$ limits the photodetector response time and as such there is a trade-off between gain and bandwidth. To minimise the transit time a high mobility channel, short electrode spacing and large electric fields are desirable. Graphene is the most promising material to achieve the unique situation in which an ultra-high carrier mobility can be accessed at the surface with micron scaled devices \cite{Banszerus2016} readily fabricated using standard electron-beam lithography techniques.

\afterpage{
	\clearpage
	\begin{landscape}
		\begin{center}
			\begin{longtable}{llcccccc}
				\caption[Graphene-based photodetectors performance]{\textbf{Summary of key performance parameters for graphene, functionalised graphene and hybrid PDs.} LDR and $D^{*}$ values are reported only if available from the experimental data. Range of $\mathfrak{R}$, $\Delta f$, $D^\star$ and LDR are given corresponding to the range in $\Delta \lambda$.} \label{tbl:SummaryPDsPerformance}\\
					
									\hline
				{\centering Ref.}  & 
				{\centering Type/Functional.}  & 
				{\centering Response} & 
				{\centering $\mathfrak{R}$ (A/W)} & 
				{\centering $\Delta f$ (Hz)}&
				{\centering $D^\star$ (Jones)}&
				{\centering $\Delta \lambda$ (nm)\textsuperscript{\emph{a}}}&
				{\centering LDR (dB)}\\
				\hline
				\multicolumn{8}{c}{Pristine graphene} \\ [5pt]
				\cite{Mueller2010} & Interdigitated & PTE & $6.1\cdot10^{-3}$ & $1.6\cdot10^{7}$ & $6\cdot10^5\,$\textsuperscript{\emph{b}} & $1500$ & $7.5$ \\
				\cite{Patil2013} & Suspended	& PTE/PV & $6.25\cdot10^{-4}$ & $10$ & $1.3\cdot10^4\,$\textsuperscript{\emph{b}} & $540$ & $24$ \\
				\cite{Lemme2011} & Dual-gated	& PTE & $1.55\cdot10^{-3}$ & $-$ & $-$ & $532$ & $-$ \\
				\cite{Mittendorff2013} & Log-antenna	& PTE & $5\cdot10^{-9}$ & $7\cdot10^{9}$ & $-$ & $30 (\mu\mathrm{m}) -220 (\mu\mathrm{m})$ & $-$ \\[5pt]
				\multicolumn{8}{c}{Functionalised graphene} \\ [5pt]
				\cite{DeSanctis2017}	& FeCl$_3$ & PV & $(0.015-0.1)\cdot10^{-3}$ & $700$ & $10^3\,$\textsuperscript{\emph{b}} & $375-10000$ & $44$ \\
				\cite{DeSanctis2017_IOP}	& FeCl$_3$ & PV & $0.1\cdot10^{-3}$ & $-$ & $-$ & $375$ & $-$ \\
				\cite{Chitara2011} & GO/rGO & PV & $0.12$ & $1.6\cdot10^{-4}$ & $-$ & $360$ & $-$\\
				\cite{Chitara2011a} & GO/rGO & PV & $4\cdot10^{-3}$ & $3\cdot10^{-2}$ & $-$ & $1550$ & $-$\\
				\cite{Yang2017} & GO/rGO & PV & $2.4\cdot10^{-4}-1.4\cdot10^{-3}$ & $2-2.5$ & $-$ & $375-118.6 (\mu\mathrm{m})$ & $7-11$\\
				\cite{Ito2016} & 3D np-rGO & PV & $1.33\cdot10^{3}-1.13\cdot10^{4}$ & $6\cdot10^{-4}$ & $-$ & $370-895$ & $4$\\
				\cite{Qi2013} & GO/Na$_2$So$_4$ & PV & $(17.5-95.8)\cdot10^{-3}$ & $2-50\cdot10^{-3}$ & $-$ & $455-980$ & $-$\\
				\cite{Lai2014} & GO & PV & $1\cdot10^{-3}-1\cdot10^{-6}$ & $2.2$ & $3\cdot10^7$ & $375-1610$ & $25$\\
				\cite{Chang-Jian2012} & GO & PV & $1.6\cdot10^{-7}-1.8\cdot10^{-6}$ & $7\cdot10^{-3}$ & $-$ & $1064$ & $-$\\
				\cite{Liu2015} & rGO/ZnO & PV & $1\cdot10^{-7}-3\cdot10^{-7}$ & $3.3$ & $-$ & $532-1064$ & $11$\\
				\cite{Chao2010} & rGO/TiO$_2$ & PV & $-$ & $0.1$ & $-$ & $>400$ & $-$\\
				\cite{Du2017} & FG & PG & $1000-10$ & $3$ & $4\cdot10^{11}-1\cdot10^{9}$ & $255-4290$ & $4$\\
				\cite{Wang2015}	& BTS/ATS SAMs & PTE & $0.02$ & $100$ & $-$ & $532$ & $15$ \\ [5pt]
				\multicolumn{8}{c}{QDs, Organics and heterostructures} \\ [5pt]
				\cite{Konstantatos2012} & PbS QDs & PG & $5\cdot10^7$ & $10$ & $7\cdot10^{13}$ & $600-1750$ & $30$\textsuperscript{\emph{b}} \\
				\cite{Sun2012d} & PbS QDs & PG & $1\cdot10^6$ & $1.2$ & $-$ & $895$ & $-$ \\
				\cite{Guo2013} & ZnO QDs & PG & $1\cdot10^4$ & $-$ & $-$ & $325$ & $-$ \\	
				\cite{Shao2015} & ZnO QDs & PG & $1\cdot10^4$ & $0.07$ & $5.1\cdot10^{13}$ & $335$ & $36$ \textsuperscript{\emph{b}} \\
				\cite{Dang2015} & ZnO QDs & PG & $2.5\cdot10^6$ & $-$ & $-$ & $326$ & $-$ \\
				\cite{Spirito2015} & CdS NPs & PG & $4\cdot10^4$ & $1000$ & $1\cdot10^{9}$ & $349$ & $-$ \\
				\cite{Robin2016} & CdSe/CdS NPs & PG & $10$ & $10$ & $10^6$ & $532-800$ & $-$\\
				\cite{Nikitskiy2016} & PbS QDs/ITO & PG/PD & $2\cdot10^6$ & $4\cdot10^{3}$ & $1\cdot10^{13}$ & $635-1600$ & $110$ \\
				\cite{Ni2017} & Si QDs & PG & $0.1 - 2\cdot10^{9}$ & $-$ & $10^3-10^{13}$ & $375-3900$ & $-$ \\
				\cite{Bessonov2017} & PbS QDs/MAPbI3 & PG & $2\cdot10^{5}$ & $100$ & $5\cdot10^{12}$ & $400-1500$ & $24$ \\
				\cite{Lee2015c} & MAPbI$_3$ & PG & $18-180$ & $4$ & $1\cdot10^9$ & $400-1000$ & $-$ \\
				\cite{Wang2015e} & MAPbBr$_2$I & PG & $6\cdot10^4$ & $2.9$ & $-$ & $405-633$ & $-$  \\
				\cite{Sun2016} & MAPbI$_3$ + Au NPs & PG & $2.1\cdot10^3$ & $0.2$ & $-$ & $532$ \\
				\cite{Chang2017} & MAPbI$_3$ & PG & $1.7\cdot10^7$ & $0.4$ & $2\cdot10^{15}$ \textsuperscript{\emph{b}} & $450-700$ & $-$ \\
				\cite{Chen2013} & Chlorophyll & PG & $1.1\cdot10^6$ & $0.78$ & $-$ & $400-700$ & $-$ \\
				\cite{Liu2014a} & Ruthenium & PG & $1\cdot10^5$ & $0.125$ & $-$ & $450$ & $-$ \\
				\cite{Huisman2015} & P3HT & PG & $1.7\cdot10^5$ & $5.8$ & $-$ & $500$ & $-$ \\
				\cite{Liu2016} & C$_8$-BTBT & PG & $1.6\cdot10^4$ & $14$ & $-$ & $355$ & $-$ \\
				\cite{Jones2017} & Rubrene & PG & $1\cdot10^7$ & $0.014$ & $9\cdot10^{11}$ & $400-600$ & $-$ \\
				\cite{Roy2013} & MoS$_2$ & PG & $5\cdot10^8$ & $-$ & $-$ & $635$ & $-$ \\
				\cite{Roy2017} & MoS$_2$ & PG & $1\cdot10^9$ & $-$ & $1\cdot10^{12}$ & $609$ & $-$ \\
				\cite{Zhang2014} & MoS$_2$ & PG & $1\cdot10^7$ & $-$ & $-$ & $650$ & $-$ \\
				\cite{DeFazio2015} & MoS$_2$ & PG & $46$ & $-$ & $-$ & $642$ & $-$ \\
				\cite{Lu2016} & GaSe & PG & $4\cdot10^5$ & $35$ & $1\cdot10^{10}$ & $532$ & $-$ \\
				\cite{Yu2017a} & MoTe$_2$ & PG & $970$ & $4.5$ & $1.6\cdot10^{11}$ & $1064$ & $-$ \\
				\cite{Mehew2017} & WS$_2$ & PG & $1\cdot10^6$ & $1500$ & $3.8\cdot10^{11}$ & $400-700$ & $12$ \\
				\cite{Liu2014}	& Tunnel barrier & PG & $1.1-10^3$ & $35$ & $-$ & $532-3200$ & $-$ \\
				\hline
			\end{longtable}
		\end{center}
			
		\textsuperscript{\emph{a}} Unless other units specified; \textsuperscript{\emph{b}} Values calculated from the published data;
	\end{landscape}
	\clearpage
}

\section{Functionalised graphene photodetectors}
In this section we will review some of the main functionalisation strategies used to enhance photo-detectivity in graphene-based devices, with particular focus to intercalation with ferric chloride (FeCl$_3$) and the use of graphene oxide (GO) as the two main forms of graphene functionalisation in which PDs with exceptional performances and scalability have been demonstrated. Other functionalised graphene PDs, such as those based on fluorographene (FG), will also be reviewed.

\subsection{FeCl$_3$-intercalated graphene photodetectors}
\begin{figure}
	\centering
	\includegraphics[width= 150 mm]{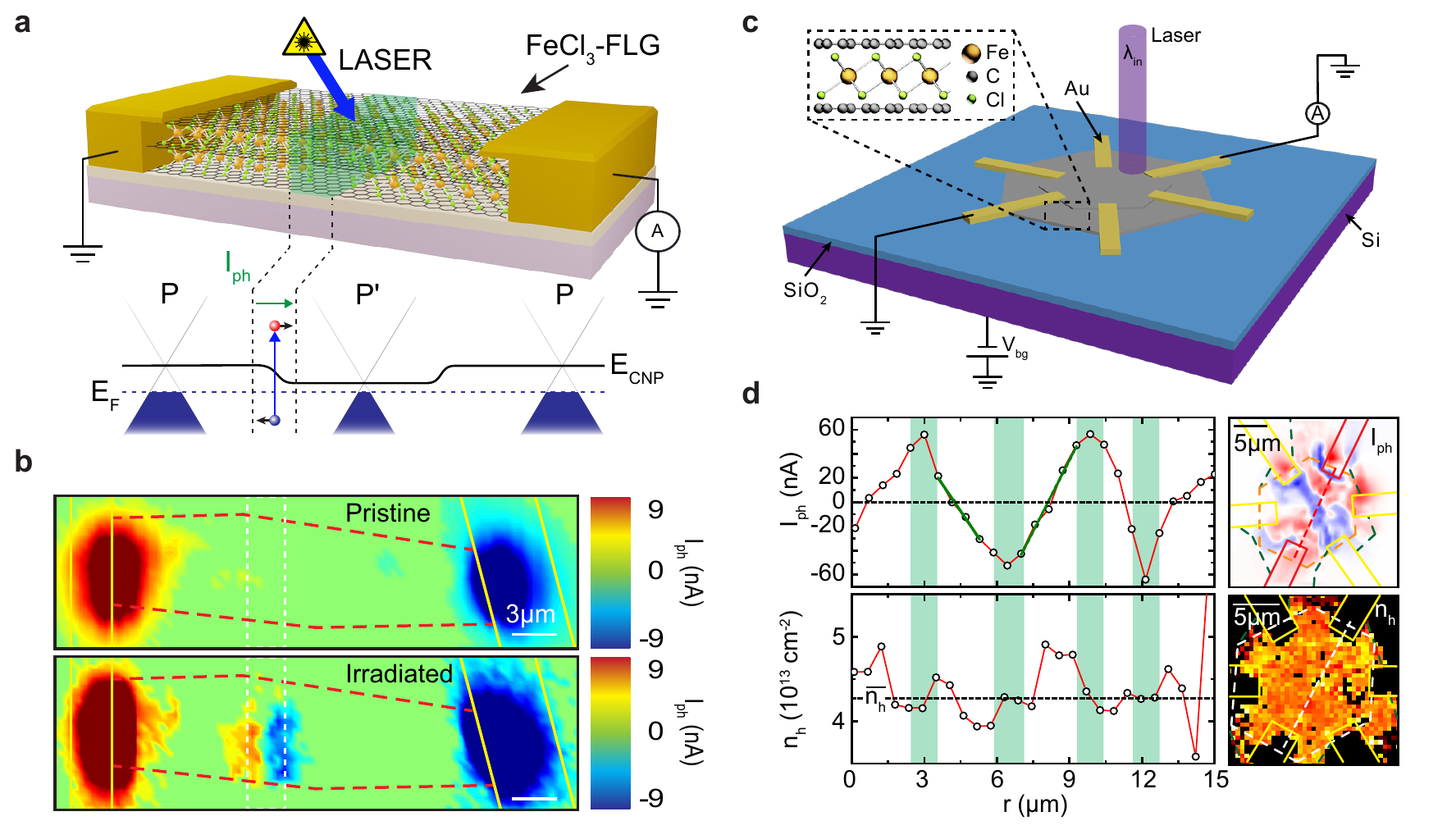}
	\caption{\textbf{FeCl$_3$-FLG Photodetectors.} (\textbf{a}) Laser-defined p-p' junction in FeCl$_3$-FLG device, $E_{CNP}$ is the charge neutrality point and $E_F$ is the Fermi level. (\textbf{b}) Scanning photocurrent ($I_{ph}$) maps before (top) and after (bottom) laser-irradiation along the white dashed lines. Red dashed lines delimit the FeCl$_3$-FLG flake, laser wavelength was $375\,\mathrm{nm}$. Adapted and reprinted with permission from De Sanctis et al \cite{DeSanctis2017}, under CC-BY license from AAAS, 2017. (\textbf{c}) Schematic of a multi-terminal hexagonal-domain FeCl$_3$-FLG photodetector and measurement circuit. (\textbf{d}) Photocurrent ($I_{ph}$, top) and charge density ($n_h$, bottom) extrapolated from the corresponding maps (right panels); Green shaded areas represent maxima and minima of the photocurrent which correspond to a change in charge density. Reproduced with permission from De Sanctis et al \cite{DeSanctis2017_IOP}, under CC-BY license from IOP Publishing Ltd, 2017.}
	\label{fig:Figure3}
\end{figure}

The intercalation of graphite with different chemical species is a well-known route to modify its bulk properties \cite{Dresselhaus1981}. More recently, few-layer graphene (FLG), i.e. between 2 and 5 layers, has been employed as a novel platform to exploit the intercalation technique. For instance, intercalation of few-layer graphene with ferric chloride (FeCl$_3$-FLG) results in a new material with enhanced optical and electrical properties \cite{Khrapach2012}. The strong charge-transfer between graphene and FeCl$_3$ molecules \cite{Zhan2010} induces large p-doping of graphene \cite{Bointon2015}, up to $10^{14}\,\mathrm{cm^{-2}}$, and drastically changes the carriers dynamics \cite{Zou2010}. The intercalation also results in a superior transparent conductor with a sheet resistance as low as $8\,\Omega/sq$ with $85\,\%$ transparency \cite{Khrapach2012} highly sought for sensing and efficient lighting technologies \cite{Alonso2016}. Contrary to bulk graphite, the intercalation of FLG takes place at relatively low temperatures using a three-zones furnace, it does not require a carrier gas and the time-scale is reduced from tens of days to only $8$ hrs. This, together with its environmental stability \cite{Wehenkel2015} and its peculiar plasmonic properties \cite{Bezares2017}, make FeCl$_3$-FLG the ideal candidate for novel opto-electronics devices \cite{Walsh2018}.

Indeed, the ability to selectively intercalate graphene led to the engineering of its photo-response. As shown in figure \ref{fig:Figure3}a, a laser beam can be used to control the microscopic arrangement of FeCl$_3$ molecules by selectively de-intercalating FeCl$_3$ from the graphene layers. In this way, a photoactive p-p' junction can be defined, as shown in the scanning-photocurrent maps (SPCM) in figure \ref{fig:Figure3}b. These laser-defined junctions have been shown to have an extraordinary LDR of $44\,\mathrm{dB}$, the highest reported in an all-graphene device (see also table \ref{tbl:SummaryPDsPerformance}). Furthermore, the spectral responsivity of such devices was also maintained, as they have been demonstrated to operate from ultra-violet (UV) to mid-infrared (MIR) wavelengths (see figure \ref{fig:Figure2}c). The key of this performance relies on the carriers dynamic engineered in these junctions. In graphene, the LDR is limited by the PTE effect and the limited density of states (DOS) available for photo-excited carriers. Since both PTE and PV effects can contribute to the observed photoresponse, careful analysis of the resulting power dependencies (see equations \ref{eq:PTEProportional} and \ref{eq:PhotocurrentPhotoCond}) and the direction of the observed photocurrent demonstrate that the in FeCl$_3$-FLG the response is dominated by PV effects, whilst PTE effects are strongly quenched \cite{DeSanctis2017}. For example, hot-carriers dynamic in graphene prevent its use in high-resolution sensing due to the smearing of the photoactive region up to several tens of microns \cite{Lemme2011}. However, by quenching such effects in laser-defined junctions in FeCl$_3$-FLG it is possible to overcome this limitation and surpass the diffraction-limit of far-field microscopy, as demonstrated using near-field techniques. In this way photoactive regions with a peak-to-peak distance of $250\,\mathrm{nm}$, (less than half the laser wavelength used) were fabricated. Such nano-scale photoactive junctions hold the promise for the realisation of novel devices in biomedical applications \cite{DeSanctis20170057}.

FeCl$_3$-FLG has also been used to realise an all-graphene position-sensitive photodetector (PSD) \cite{DeSanctis2017_IOP}. In this case a CVD-grown hexagonal domain of bilayer graphene \cite{Mohsin2013} was used as starting material and intercalated following the same procedure used in previous works \cite{Khrapach2012,Bointon2015,Bointon2014}. A multi-terminal geometry, arranged along the edges of the hexagons, allows to measure photocurrent between different pairs of opposing contacts (see figure \ref{fig:Figure3}c). Strikingly, the observed photocurrent displayed a bipolar and monotonic behaviour in regions in which the intercalation-induced charge density changes abruptly. These changes were shown to be related to the partial intercalation of the FLG along specific lines irradiating from the centre of the hexagons (figure \ref{fig:Figure3}d). Therefore, photoactive p-p' junctions were formed in the hexagonal crystal at the growth stage. This work was the first demonstration of position-sensitive behaviour in an all-graphene PD (i.e. where both the active and transport layers are made of graphene).

The responsivity of FeCl$_3$-FLG PDs is in line with other pristine graphene devices ($0.1-1\,\mathrm{mA/W}$) and it is limited by the absence of a gain mechanism able to multiply the photo-generated carriers in the material. However, the suppression of PTE increases the operating bandwidth of these detectors up to $700\,\mathrm{Hz}$, when compared to other functionalised graphene PDs (such as GO) this is two to four orders of magnitude higher. In FeCl$_3$-FLG PDs the speed is limited by the diffusion time of the excited carriers which is affected by the reduced mobility due to the high levels of doping \cite{Banerjee2016,Khrapach2012,Bointon2015}.

\subsection{Graphene oxide}
A functionalised form of graphene decorated with oxygen atoms (in the form of carboxyl, hydroxyl or epoxy groups) is graphene oxide (GO) \cite{Mkhoyan2009, Dreyer2010}. GO is an insulator in which the energy bandgap and electrical conductivity can be tuned by reducing its oxygen content in what is known as reduced graphene oxide (rGO) \cite{Compton2010}. GO is usually prepared in solution and it is a scalable way of obtaining graphene-based materials \cite{Dikin2007}, composites \cite{Stankovich2006} and devices for microelectronics \cite{Eda2008}. Further functionalisation is possible in GO and rGO by replacing oxygen groups with other chemical species \cite{Qi2013} or by interfacing GO with other nano-structured materials such as nanoparticles \cite{Chao2010}.

\begin{figure}
	\centering
	\includegraphics[width= 150 mm]{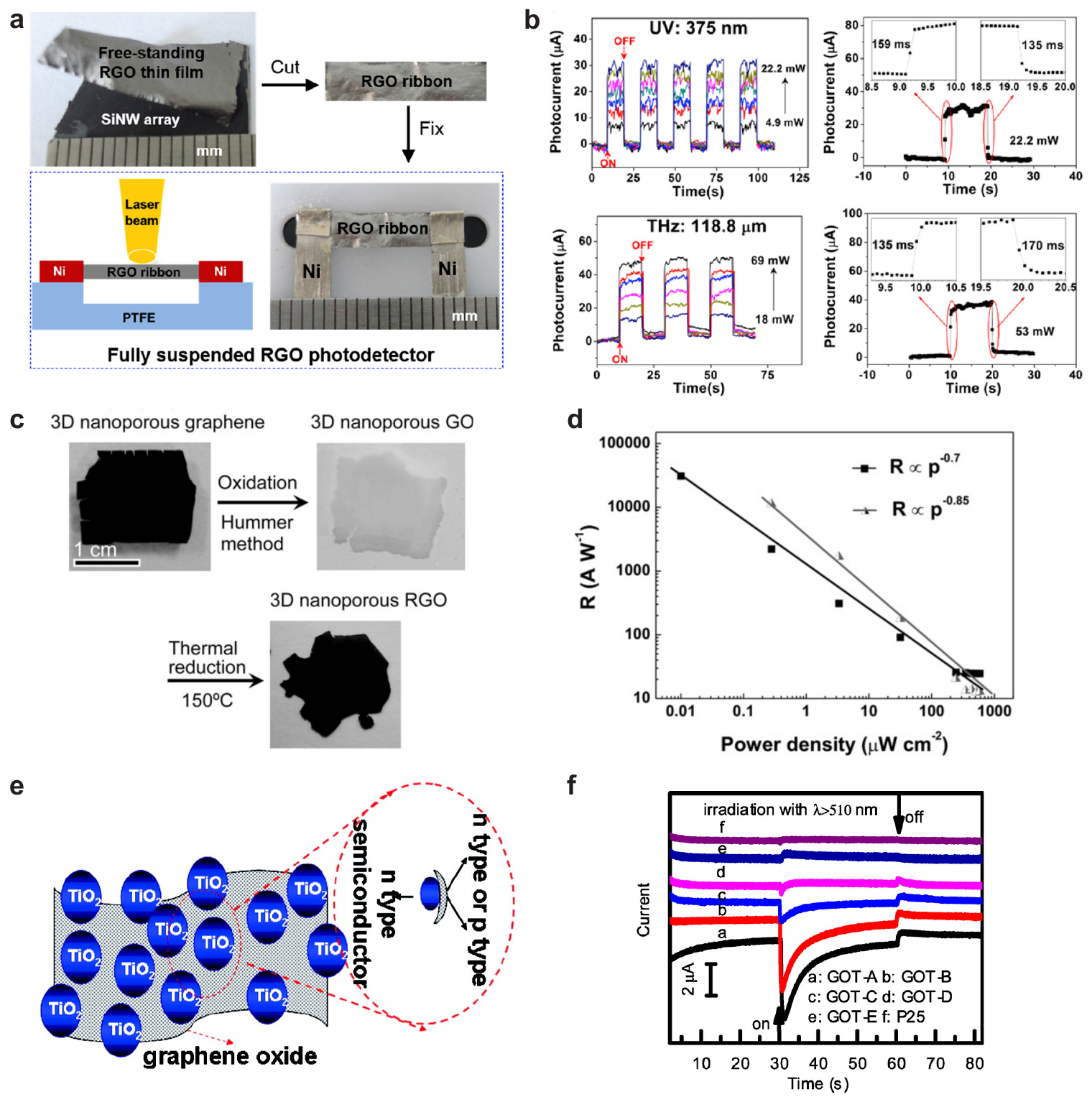}
	\caption{\textbf{GO and rGO Photodetectors.} (\textbf{a}) Fully suspended rGO photodetector grown on Si nanowire array and (\textbf{b}) photoresponse of this photodetector under UV and THz illumination. Reprinted with permission from Yang \textit{et al} \cite{Yang2017}, \textit{Carbon} 115, 561-570, Copyright 2017 Elsevier. (\textbf{c}) Fabrication steps of 3D nanoporous rGO (3D np-rGO). (\textbf{d}) Power-dependent responsivity for two sample 3D np-rGO devices at $\lambda = 370\,\mathrm{nm}$. Reprinted with permission from Ito \textit{et al} \cite{Ito2016}, \textit{Adv. Func. Mater.} 26, 1271. Copyright 2016 John Wiley and Sons. (\textbf{e}) Schematic of TiO$_2$ nanoparticles embedded in GO matrix. (\textbf{f}) Cathodic photoresponse of GO/TiO$_2$ photodetectors under visible illumination, on and off indicate the switching of the light source. Reproduced with permission from Chen C. \textit{et al} \cite{Chao2010}, \textit{ACS Nano} 2010 4, 6425. Copyright 2010 American Chemical Society.}
	\label{fig:Figure4}
\end{figure}

The photoresponse of GO has been investigated by many authors in devices with different degrees of reduction and functionalisation. Solution-processed GO has been used to demonstrate photodetection across a wide spectral range (see figure \ref{fig:Figure2}c). Chitara \textit{et al} \cite{Chitara2011} demonstrated that chemically-reduced GO solution \cite{Park2009} can be used to realise a UV-sensitive photodetector with a responsivity of $120\,\mathrm{mA/W}$ operating at $\lambda = 360\,\mathrm{nm}$. Soon after, the same authors demonstrated that a similar photodetector can operate at infra-red (IR) wavelengths ($\lambda = 1550\,\mathrm{nm}$) \cite{Chitara2011a} although with a reduced responsivity of $4\,\mathrm{mA/W}$. In both cases the operating bandwidth of these devices was of the order of $0.1-30\,\mathrm{mHz}$, corresponding to rise and fall times of several seconds. The works of Chitara \textit{et al} show that GO PDs can operate across an exceptionally large range of wavelengths. Recently, Yang \textit{et al} \cite{Yang2017} demonstrated a free-standing rGO photodetector able to operate from UV ($\lambda = 375\,\mathrm{nm}$) to THz ($\lambda = 118.6\,\mu\mathrm{m}$) wavelengths (see figure \ref{fig:Figure4}a-b) with responsivity ranging between $0.24\,\mathrm{mA/W}$ and $1.4\,\mathrm{mA/W}$ (see table \ref{tbl:SummaryPDsPerformance}) and an operating bandwidth of $2-2.5\,\mathrm{Hz}$. This technology is characterized by a LDR in the range $7-11\,\mathrm{dB}$. The ease of fabrication and the scalability of the material make this kind of suspended rGO photodetectors very attractive for macroelectronics. Other authors compared the photoresponse of GO and rGO photodetectors. Chang-Jian \textit{et al} \cite{Chang-Jian2012} prepared GO and rGO photodetectors form GO solution and demonstrated that photocurrent from rGO devices was due to the separation of excited electron-hole pairs (PV-type of response), where electrons are injected into the contact at higher potential (positive), thus giving an increase in photocurrent. On the contrary, in GO devices a cathodic photocurrent was observed, which can be attributed to the injection of holes into the negative contact due to the work-function mismatch between the GO and the Au contacts. Such detectors, when illuminated with IR light ($\lambda = 1064\,\mathrm{nm}$) showed responsivity of $1.8\,\mu\mathrm{A/W}$ and operating bandwidth of $7\,\mathrm{mHz}$.

Qi \textit{et al} investigated the response of solution-processed GO in the Visible-IR range. By drop-casting GO solution onto a glassy carbon electrode they were able to measure the photoresponse of this material in Na$_2$SO$_4$ solution. Their experiments report responsivity values of $95.8-17.5\,\mathrm{mA/W}$ for $\lambda = 455-980\,\mathrm{nm}$ with an operating bandwidth of $2-50\,\mathrm{mHz}$. In the same work they demonstrate UV sensitivity under $\lambda = 280-350\,\mathrm{nm}$ illumination, although the data doesn't allow the estimation of the responsivity at this wavelengths. The solution-processed approach is very scalable and attractive for applications in the chemical industry where PDs can be used to work in specific solutions. However, these detectors were found to be unstable in ambient conditions and degraded under UV illumination.

Low responsivity in GO devices is attributed to the poor electrical contact between stacked flakes, which is a result of solution-processing techniques, and the reduced light absorption due to the 2D nature of the individual flakes. Furthermore, the large amount of defects in GO and rGO films limits the charge carriers mobility and increases charge recombination sites, limiting the overall efficiency of the device. In order to improve the responsivity of GO photodetectors, Ito \textit{et al} \cite{Ito2016} developed a device based on 3D nanoporous rGO (3D np-rGO, see figure \ref{fig:Figure4}c). This material displays and enhanced absorption in the near-UV region ($\lambda = 300-400\,\mathrm{nm}$) indicating a large density of states at high photon energies. However, the long absorption tail indicates that defect states allow photon absorption across the UV-Visible and IR range. Indeed, the photoresponse of np-rGO was demonstrated to strongly depend on the reduction time and to reach a maximum responsivity of $1.13\times10^4\,\mathrm{A/W}$ at $\lambda = 370\,\mathrm{nm}$ and $1.33\times10^3\,\mathrm{A/W}$ at $\lambda = 895\,\mathrm{nm}$ for samples reduced for $150\,\mathrm{min}$. However, as it can be seen in  the power-dependence of the responsivity in the UV range (see figure \ref{fig:Figure4}d), such high values of responsivity are measured at very low power densities ($0.01\,\mu\mathrm{Wcm^{-2}}$) and rapidly decreases by more than three orders of magnitude as the power density reaches $300\,\mu\mathrm{Wcm^{-2}}$. Furthermore, the saturation of the responsivity at higher power suggests that the device is operating above the NEP on this region only, giving an estimate for the LDR of $\sim4\,\mathrm{dB}$. The behaviour of the power-dependent responsivity suggests that the photocurrent mechanism is that of a photoconductor, whereby illumination creates a population of electron-hole pairs which are separated by an applied electric field across the semiconductor. The operating bandwidth can be estimated from the measured decay time and equals $6\,\mathrm{mHz}$, in line with other GO-based photodetectors. 

In order to improve the performance of GO PDs and to add functionalities such as photocatalytic properties, several groups worked on preparing composite materials combining solution-processed GO and rGO with oxide nanoparticles such as TiO$_2$ or ZnO. Chen \textit{et al} \cite{Chao2010} demonstrated a GO/TiO$_2$ composite which can be used as a photodetector (see figure \ref{fig:Figure4}e). In their experiments they found that both a cathodic and anodic photoresponse was possible depending on the starting concentration of GO in the initial solution (see figure \ref{fig:Figure4}f). This indicates a different doping of the GO composite depending on the amount of TiO$_2$ in the initial solution (see figure \ref{fig:Figure4}e). The photoresponse was observed at wavelengths $>400\,\mathrm{nm}$ across the Visible-NIR spectrum. Interestingly, this composite material displayed photocatalytic properties when illuminated with wavelengths $>400\,\mathrm{nm}$, as shown by the degradation of methyl orange. Liu \textit{et al} \cite{Liu2015} developed a rGO/ZnO nanowire PD. In this case, the hybrid PD behaves like a photodiode, rather than a photoconductor. The responsivity of this device is in line with others in the range $0.1-0.3\,\mu\mathrm{A/W}$ and it is able to operate across the visible-IR spectrum ($\lambda = 532-1064\,\mathrm{nm}$) with a LDR of $25\,\mathrm{dB}$ and an operating bandwidth of $6\,\mathrm{mHz}$. The PV response of this hybrid PD was demonstrated to be due to the Schottky junction between the ZnO nanowires and the GO, which also accounts for the relatively low responsivity.

The pursuit of an environmentally friendly method to produce electronic materials led to research into ways to make graphene and GO from many sources. Lai \textit{et al} \cite{Lai2014} demonstrated a vertical junction photodetector made from GO produced by processing of glucose solution \cite{Tang2012}. This device structure comprises two layers of GO prepared with different annealing temperatures sandwiched between an ITO and a Au electrode. Such device is shown to operate from deep-UV ($\lambda = 290\,\mathrm{nm}$) to IR ($\lambda = 1610\,\mathrm{nm}$) across the whole visible spectrum. They report responsivity values between $1\,\mathrm{mA/W}$ and $1\,\mu\mathrm{A/W}$ going from UV to IR with a LDR of $25\,\mathrm{dB}$ at ($\lambda = 410\,\mathrm{nm}$). The operating bandwidth of this device (see table \ref{tbl:SummaryPDsPerformance}) is in the range $1-6\,\mathrm{Hz}$, in line with other reports on GO PDs. The photoconductive behaviour of this PD and the linearity of the power-dependent photocurrent (see equation \ref{eq:PhotocurrentPhotoCond}) allow to conclude that the PV photocurrent generation is responsible for its photo-activity.

From the data reported in literature, it is clear that all GO-based PDs have a very slow response time. This can be attributed to the intrinsic defective nature of the active material. Charge traps in the GO act as pinning centres for photoexcited electron-hole pairs and allow both fast recombination, which limits the responsivity, and slow carrier drift/diffusion, which limits the speed \cite{Joung2010}. However, the presence of these defects states and the gapless nature of rGO allow to absorb light across a wide spectral range, spanning from deep-UV to THz (see figure \ref{fig:Figure2}), albeit with large variations in responsivity. In particular, the large sensitivity in the UV region makes GO and rGO PDs extremely promising for replacing current semiconductors in applications such as environmental monitoring, water purification and defence \cite{Sang2013}.

\subsection{Other functionalised graphene PDs}
\begin{figure}
	\centering
	\includegraphics[width= 150 mm]{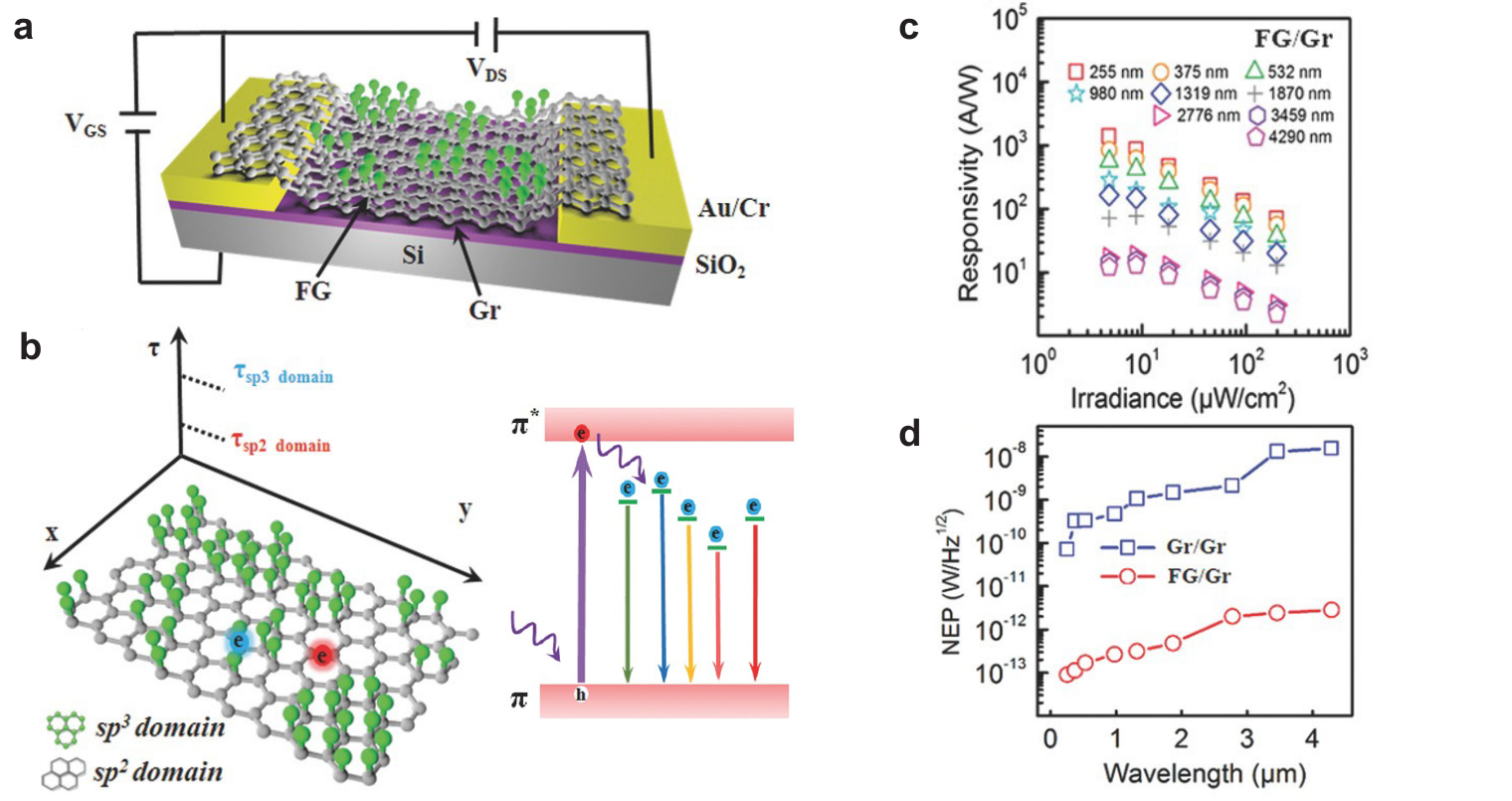}
	\caption{\textbf{FG Photodetectors.} (\textbf{a}) Device schematic of a FG/graphene PD. (\textbf{b}) Domain structure of partially-fluorinated graphene and corresponding lifetimes of photoexcited charges (left). Trapping and relaxation schematic in FG responsible for the observed photo-response. (\textbf{c}) Responsivity as a function of incident power density for multiple wavelengths and (\textbf{d}) comparison of NEP as a function of incident wavelength for a graphene/graphene and a FG/graphene device. Reprinted with permission from Du \textit{et al} \cite{Du2017}, \textit{Adv. Mater.} 29, 1700463. Copyright 2017 John Wiley and Sons.}
	\label{fig:Figure5}
\end{figure}

Similar to GO, the attachment of fluorine atoms to graphene results in another insulating form of functionalised graphene, known as fluorographene (FG) \cite{Withers2011a}. As with rGO, the degree of fluorination in FG can be changed during fabrication \cite{Cheng2010} or afterwards using, for example, e-beam irradiation \cite{Withers2011,Martins2013}. The ability to selectively tune the insulating properties of FG gives a versatile material for photodetector technology. Du \textit{et al} \cite{Du2017} investigated a FG/graphene photodetector where graphene is used as a charge transport layer and FG as the charge-trapping layer, creating a photogating-based (PG) device (figure \ref{fig:Figure5}a). Fluorination modifies the C-C bond hybridisation from $\mathrm{sp}^2$ to $\mathrm{sp}^3$. These confined areas have different charge trapping times, while the pristine graphene layer acts as high-mobility layer, enabling charge re-circulation and high gain (see equation \ref{eq:resp} and figure \ref{fig:Figure5}b). This photodetector works over a broad range of wavelengths, from UV ($\lambda = 255\,\mathrm{nm}$) to MIR ($\lambda = 4.29\,\mu\mathrm{m}$), albeit MIR range was tested only at a temperature of $77\,\mathrm{K}$. Responsivity varies across the spectrum, from a maximum of $10^{3}\,\mathrm{A/W}$ in the UV to $10\,\mathrm{A/W}$ in the MIR range, and with incident optical power (see figure \ref{fig:Figure5}c). Indeed, a drop in the responsivity as a function of power indicates saturation of the photocurrent, as shown in figure \ref{fig:Figure5}c. From the data provided (figure \ref{fig:Figure5}d) we can estimate a LDR of $4\,\mathrm{dB}$ and an operating bandwidth of $3\,\mathrm{Hz}$, in line with the high responsivity values and the slow response given by the trap states in the FG. Overall, the performance of the device varies with the degree of fluorination (i.e. C/F ratio)  and it is observed to be maximum for all wavelengths for a C/F ratio of $3.5-3.75$. Indeed, another study highlighted the role of the density of defect states in the photoresponse of FG/graphene photodetectors \cite{Mehew2017SPIE}. Statistical analysis of several samples also shows a small sensitivity to fabrication variables since the responsivity is observed to vary within a factor of $1.5$ at maximum. Interestingly, given the different trapping time in the $\mathrm{sp}^2$ and $\mathrm{sp}^3$ regions, the authors demonstrate that it is possible to extract the different contributions using AC+DC modulation of the incident light, where the DC contribution to the photocurrent is attributed to the $\mathrm{sp}^3$ sites (slow traps) and the AC response to the $\mathrm{sp}^2$ ones (fast traps). The ratio between the charge trapping times allows to estimate  a photoconductive gain of $2\times10^5$, weakly dependent on the wavelength.

A different approach to graphene functionalisation was followed by Wang \textit{et al} \cite{Wang2015}. In this work they present a p-n junction graphene PD realised using CVD graphene on top of silane-modified SiO$_2$. Self-assembled monolayers (SAMs) of 3-aminopropyltriethoxysilane (ATS) and N-butyltriethoxysilane (BTS) where put in contact with graphene to achieve p- and n- doping in adjacent regions. SPCM revealed photocurrent generation at the interface between these two regions. Responsivity of the order of $0.02\,\mathrm{mA/W}$ and an operating bandwidth of $100\,\mathrm{Hz}$ were achieved in the visible range. The photoresponse of the device was measured to be due to PTE, enabled by the difference in Seebeck coefficient between the n- and p- region. Power-dependent photocurrent allows to estimate a LDR of $15\,\mathrm{dB}$ for this device. The use of SAMs and CVD graphene makes this technique scalable and the device performance is promising for sensing and imaging applications. However, the environmental stability and bio-compatibility of the silane-based SAMs have not been tested.

\section{Hybrid and heterostructure photodetectors}
The inherent low responsivity of graphene-based devices is related to its small absorption per layer and to the lack of a mechanism able to  multiply the photogenerated carriers. Therefore the maximum quantum efficiency of a pure graphene device cannot exceed $1$. Such limit can be overcome if a gain mechanism is present (equation \ref{eq:resp}). One way to attain this is to combine graphene with a photoactive material and use the high mobility of the graphene channel to extract one photoexcited carrier, enabling charge recirculation. We will consider different light-sensitizing materials used in graphene-based PDs, including quantum dots (QDs), perovskites, organics and TMDs.

\subsection{Graphene/Quantum Dots and perovskites interfaces}
Quantum Dots (QDs) are highly suited as a light-sensitizing material for graphene due to strong absorption from UV to NIR, size tunable band gap and low temperature deposition \cite{Konstantatos2006}. Often these are synthesised in solution and spin coated onto a target substrate. Such processing means that production can readily be scaled up and non-destructively applied \cite{Klekachev2011}. This is particularly important for graphene as conventional material deposition processes are known  to induce defects and disorder \cite{Wang2012b}.

\begin{figure}
	\centering
	\includegraphics[width=150 mm]{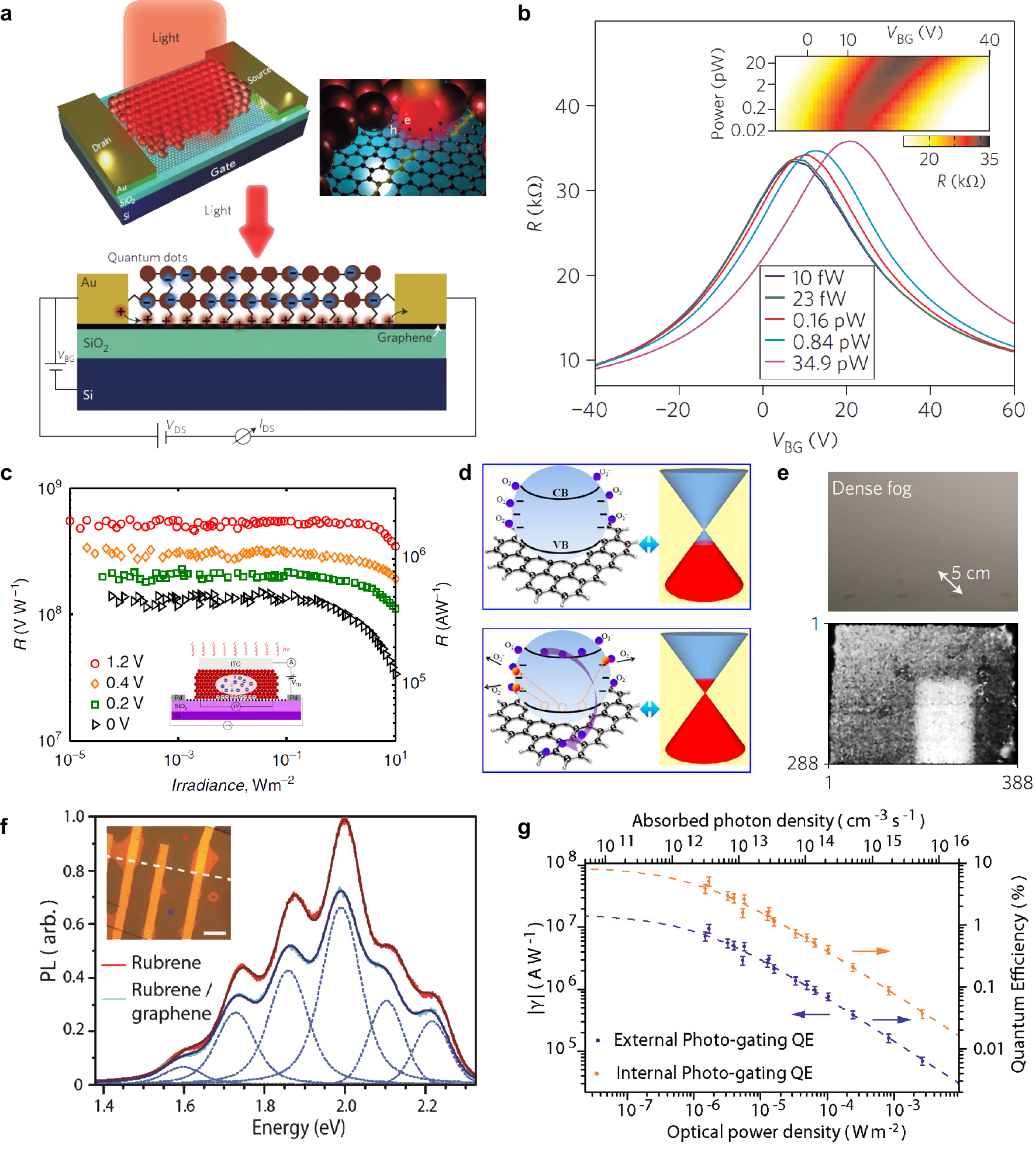}
	\caption{\textbf{Quantum Dots and Organic graphene hybrids.} (\textbf{a}) Schematic of PbS QD sensitised graphene channel with electrical connections. (\textbf{b}) Back-gate dependence of graphene channel resistance under different illumination powers demonstrating photogating effect. Inset shows two-dimensional plot of power and back-gate voltage. Reproduced with permission from Konstantatos \textit{et al} \cite{Konstantatos2012} Copyright 2012, Nature Publishing Group. (\textbf{c}) Plateau in responsivity vs optical power shows the extended LDR of ITO-QD-graphene phototransistor. Reproduced under CC-BY 4.0 from Nikitskiy \textit{et al} \cite{Nikitskiy2016}, 2016. (\textbf{d}) Oxygen assisted charge-trapping process in ZnO QDs. Reproduced with permission from Shao \textit{et al} \cite{Shao2015} Copyright 2015 American Chemical Society. (\textbf{e}) Photos taken in dense fog of a cylindrical object by a conventional silicon (upper) and PbS-graphene hybrid (lower) sensor array. Reproduced with permission from Goossens \textit{et al} \cite{Goossens2017} Copyright 2017, Nature Publishing Group. (f) Photoluminescence of rubrene single crystal with (blue) and without (red) graphene underlayer. Inset shows optical image of device where coloured dots highlight position where the PL spectra were acquired. Reproduced with permission under CC-BY 4.0 from Jones \textit{et al} \cite{Jones2017}, 2017.
		\label{fig:GFET_NPs}
	}
\end{figure}

The first report of a hybrid graphene-quantum dots photodetector is the work of Konstanatos \textit{et al} \cite{Konstantatos2012}. A two terminal graphene FET was spin coated with an $80\,\mathrm{nm}$ thick film of lead sulphide (PbS) QDs with the first exciton peak at $950\,\mathrm{nm}$ or $1450\,\mathrm{nm}$, see figure \ref{fig:GFET_NPs}a. Photogenerated holes transfer to graphene whilst electrons remain in the QDs. A built-in field at the graphene/QD interface and charge traps within the PbS QDs keep the electron trapped for a time-scale $\tau$. The transferred holes changes the conductivity of the graphene channel leading to a shift in the Dirac point voltage, figure \ref{fig:GFET_NPs}b. Owing to the high-mobility of graphene the hole can be recirculated multiple times resulting in a gain that exceeds $10^8$ electrons/photon, yielding a responsivity of $\sim10^7\,\mathrm{A/W}$. The charge trapping mechanism limits the bandwidth to $10\,\mathrm{Hz}$, however recombination can be accelerated by providing an electrical pulse to the gate. This lowers the built-in field allowing the trapped electron to recombine with the hole. This strategy reduces the temporal response to $\sim10\,\mathrm{ms}$. Similar device performance was achieved using CVD graphene which represents an important step in the future commercialisation of such detectors \cite{Sun2012d}. To further increase the response time Nikitskiy \textit{et al} deposited an ITO top contact on the PbS-QD film \cite{Nikitskiy2016}, thereby incorporating a photodiode into the phototransistor structure. This transformed the passive sensitising layer into an active one increasing the charge collection efficiency close to $100\%$ when surface reflections are taken into account ($>70\%$ without). Applying a small voltage $\sim2\,\mathrm{V}$ to the ITO increases the depletion region at the graphene-QD interface increasing the contribution to charge collection from carrier drift over diffusion. They reported two major changes in device performance compared to the passive example \cite{Konstantatos2012}. The bandwidth is increased to $1.5\,\mathrm{kHz}$ as the carrier lifetime is now limited by the time constant of the ITO-QDs-graphene photodiode as opposed to the charge trapping lifetime. A significant enhancement in LDR to $110\,\mathrm{dB}$ was reported, representing the highest value achieved in hybrid graphene photodetectors, figure \ref{fig:GFET_NPs}d.

Other QDs have been used including Zinc Oxide (ZnO) which has a much stronger absorption in the UV than visible \cite{Guo2013,Shao2015}. Shao \textit{et al} demonstrated a UV phototransitor by coating graphene with ZnO QDs \cite{Shao2015}. A self-assembled monolayer (SAM) was deposited on the SiO$_2$/Si substrate prior to the graphene transfer resulting in an increased mobility due to a reduction in charged impurity scattering. They reported a gate-tunable responsivity of $4\times10^9\,\mathrm{A/W}$ (at $\lambda
= 335\,\mathrm{nm}$) for the device with a UV-Visible rejection ratio of $\sim10^3$. The response time exceeds $2\,\mathrm{s}$ which is likely due to the oxygen mediated charge-trapping mechanism (figure \ref{fig:GFET_NPs}c) \cite{Guo2013,Shao2015}. Ni \textit{et al} used the localized surface plasmon resonance of doped Silicon QDs to enhance the MIR absorption of a graphene phototransistor \cite{Ni2017}.

Robin \textit{et al} \cite{Robin2016} have engineered a photodetector based on the charge-transfer at the interface between graphene and CdSe NPs (CdSe-np). Such NPs, in their pristine form, display sharp excitonic peaks which makes them ideal for wavelength-selective PDs. In their work they use epitaxially-grown graphene on silicon carbide (SiC) onto which a colloidal solution of CdSe-np is drop-casted and then embedded into an ionic polymer gate. Photogating (see figure \ref{fig:Figure1}c, right panel) is the key mechanism which is responsible for photo-activity in this device: excitons created in the nanoparticles are split and one of the carriers is injected into the graphene depending on the local doping, whilst the other remains trapped in the CdSe-np. This process gives an enhanced responsivity if compared to pristine graphene of $10\,\mathrm{A/W}$ with an operating bandwidth of $10\,\mathrm{Hz}$. Furthermore, the broadband absorption of graphene combined with the excitonic features of the CdSe-np allow this device to operate across the visible and NIR range ($\lambda = 532-800\,\mathrm{nm}$).

Asides from QDs, perovskites have shown promise in the next generation of hybrid graphene photodetectors with ideal properties such as direct band gap, large absorption and relatively high mobility \cite{Lee2015c,Wang2015e,Sun2016,Chang2017}. However, the reported values for responsivity range from $10^2$ to $10^7\,\mathrm{A/W}$ with the spread in values likely due to the poor coverage and uniformity of the perovskite films. In addition perovskites are highly unstable in environments containing oxygen and water.

Recently the first image sensors based on graphene photodetectors have been reported \cite{Goossens2017,Bessonov2017}. These combine sensitised graphene with a complementary metal-oxide-semiconductor (CMOS) readout circuit through back-end-of-line integration. This results in a sensor capable of broadband light-detection (UV-NIR) - previously unobtainable in a monolithic CMOS detectors. One particular advantage is shown in figure \ref{fig:GFET_NPs}e, where the graphene sensor reveals an object shrouded by fog that would otherwise be invisible to a conventional Si-CMOS sensor.

\subsection{Graphene/organics interfaces}
Organics have also been investigated as light-sensitising elements in graphene FETs. They are of particular interest owing to their intrinsic affinity to biological systems as well as the ability to tailor their spectral selectivity through chemical functionalisation. A number of different materials have been investigated including chlorophyll \cite{Chen2013}, ruthenium \cite{Liu2014a}, P3HT \cite{Huisman2015}, and C8-BTBT \cite{Liu2016}. However the reported quantum efficiencies are inferior to those of QDs ($\sim25\%$) \cite{Konstantatos2012}. Disorder in the crystal structure could play a major role as evidenced by amorphous P3HT exhibiting an efficiency $\sim0.002\%$ \cite{Huisman2015} whilst for polycrystalline C8-BTBT $0.6\%$ \cite{Liu2016} was achieved. In the work of Jones \textit{et al} single crystals of rubrene were laminated onto CVD graphene channels \cite{Jones2017}. The long-range herringbone stacking of rubrene molecules in a single crystal allows exciton diffusion over several micrometers with rubrene exhibiting strong photoluminesence (PL) in the visible. However when placed in contact with graphene the PL intensity is suppressed by $\sim25\%$, see figure \ref{fig:GFET_NPs}f. This suggests that a large number of excitons dissociate at the graphene-rubrene interface. Indeed this hybrid photodetector exhibits a responsivity as large as $10^7\,\mathrm{A/W}$ and an internal efficiency above $5\%$ thanks to this charge transfer and charge recirculation, as shown in figure \ref{fig:GFET_NPs}g.

\subsection{Graphene/TMDs vdW Heterostructures}
\begin{figure}
	\centering
	\includegraphics[width=150 mm]{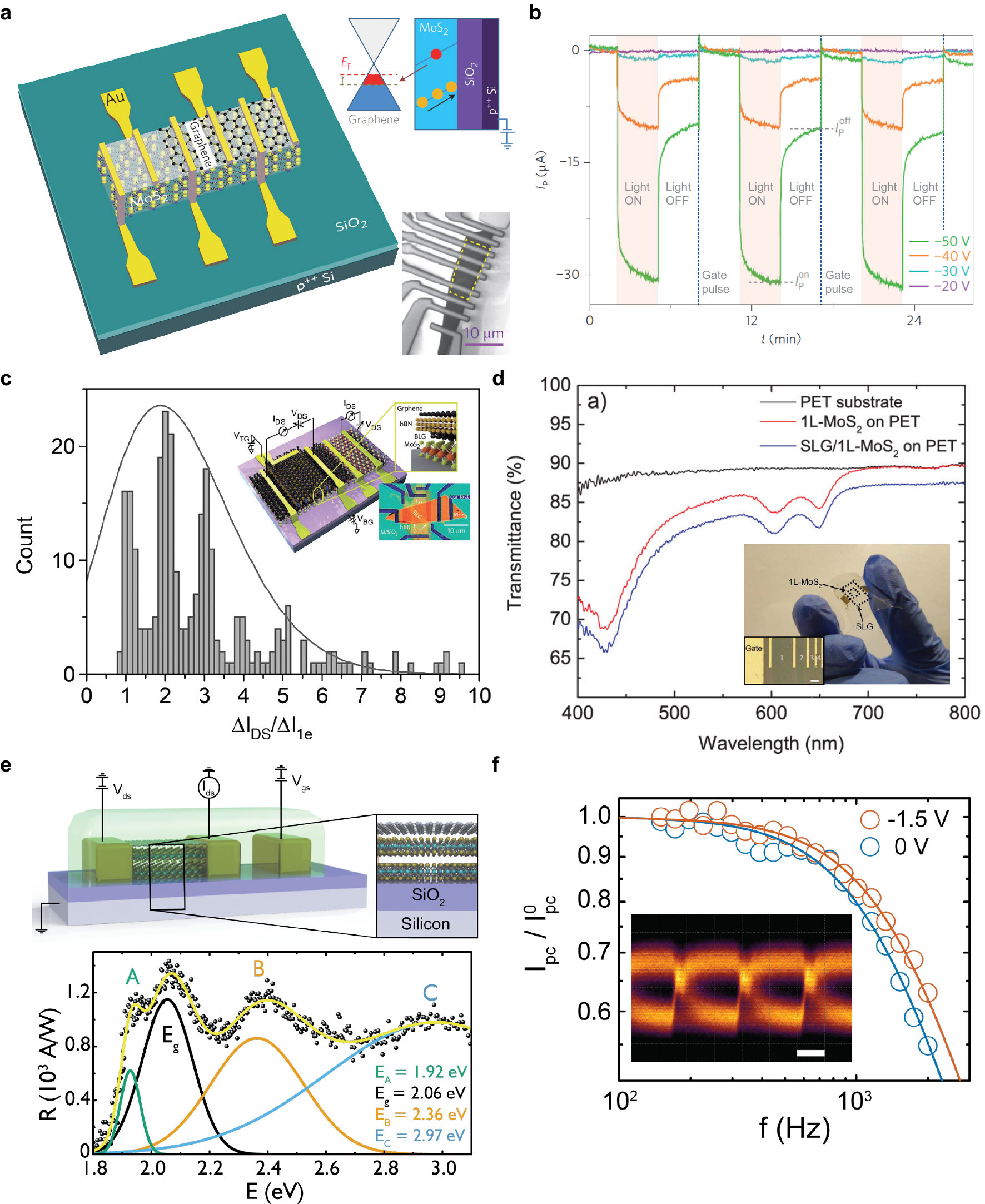}
	\caption{\textbf{TMD-graphene hybrids.} (\textbf{a}) Schematic of MoS$_2$-graphene photodetector with scanning electron microscope (SEM) image of device (lower inset) and principle of operation (upper inset). (\textbf{b}) Temporal response of photocurrent from device in (\textbf{a}) under illumination and gate pulse cycles. Reproduced with permission from Roy \textit{et al} \cite{Roy2013} Copyright 2013, Nature Publishing Group. (\textbf{c}) Histogram of normalized current shift indicating number resolved photon counting. Inset shows schematic and SEM image of device.  Reproduced with permission from Roy \textit{et al} \cite{Roy2017} Copyright 2017, John Wiley \& Sons. (\textbf{d}) Transmittance of all-CVD MoS$_2$/graphene photodetector. Optical image highlights transparency (inset). Reproduced with permission under CC-BY 4.0 from De Fazio \textit{et al} \cite{DeFazio2015}, 2015 ACS Publications. (\textbf{e}) Schematic of ionic polymer gated WS$_2$-graphene photodetector (upper) and spectral responsivity (lower). (\textbf{f}) Optical bandwidth of device in (\textbf{e}) extends to $1.5\,\mathrm{kHz}$. Inset shows eye diagram obtained at $2.9\,\mathrm{kbit/s}$. Reproduced with permission under CC-BY 4.0 from Mehew \textit{et al} \cite{Mehew2017}, 2017 John Wiley \& Sons.
	}
	\label{fig:GFET_TMDs}
\end{figure}

The family of 2D materials extends beyond graphene to include insulators (hBN), semiconductors (e.g. MoS2 and WS2), and materials exhibiting more exotic properties such as superconductivity and magnetism \cite{Novoselov2016}. Similar to graphene adjacent layers are separated by a van der Waals (vdW) gap facilitating the rapid prototyping of devices by mechanical exfoliation of bulks flakes. Recently techniques have been developed to create complex heterostructures though layer-by-layer assembly of 2D flakes \cite{Geim2013}. Naturally, the semiconducting 2D materials present an ideal sensitising layer in graphene photodetectors due to their strong-light matter absorption and visible to NIR band gap \cite{Mak2016a}.

The first demonstration of a graphene-TMD phototransistor was found in the work of Roy \textit{et al} \cite{Roy2013}. Here, graphene was placed onto a few-layer flake of molybdenum disulphide (MoS$_2$), see figure \ref{fig:GFET_TMDs}a. Electron-hole pairs are photoexcited within the MoS$_2$ and one type of charge carrier transferred to the graphene channel. Due to the formation of localized states at the MoS$_2$/SiO$_2$ interface persistent photocurrent is observed (no return to initial dark state) indicating an extremely long trapped carrier lifetime. With the ultra-high mobility of the graphene channel ($\mu=10^4\,\mathrm{cm^2V^{-1}s^{-1}}$ this translates to a gain of $5\times10^{10}$ electrons/photon and responsivity of $5\times10^8\,\mathrm{A/W}$. Using this structure the authors demonstrated a multilevel optoelectronic memory using light and gate pulses to write and erase each bit respectively, figure \ref{fig:GFET_TMDs}b. In a follow up work they replaced monolayer graphene with bilayer and by dual gating they electrostatically opened a band gap of $\sim90\,\mathrm{meV}$ \cite{Roy2017}, which resulted in a reduction in the channel noise by $6$ to $8$ orders of magnitude. This allowed them to demonstrate a number-resolved photo counter capable of determining the poissonian emission statistic of an LED, figure \ref{fig:GFET_TMDs}c. Several groups have extended these prototypes to all large-area CVD versions \cite{Zhang2014,DeFazio2015}, without a loss of the significant gain mechanism. De Fazio \textit{et al} used monolayers of graphene and MoS$_2$ to realise a semitransparent ($92\%$ transmittance at $\lambda = 642\,\mathrm{nm}$) and flexible photodetector suitable for wearable applications, figure \ref{fig:GFET_TMDs}d. To extend the spectral range into the NIR different 2D materials have been used as the photoactive material including Bi$_2$Te$_3$ ($E_g \sim 0.3\,\mathrm{eV}$) \cite{Qiao2015} and MoTe$_2$ ($E_g \sim 1.1\,\mathrm{eV}$) \cite{Yu2017a}. However, in these cases the responsivity is much lower than reported for MoS$_2$, typically less than $10^3\,\mathrm{A/W}$. 

The slow temporal response of these heterostructure photodetectors limits their use to steady state imaging applications as the presence of long-lived charge traps that provide the extreme photo-sensitivity comes at the expense of operational speed. Recently, Mehew \textit{et al} fabricated graphene–tungsten disulphide (WS$_2$) heterostructure photodetectors encapsulated in the ionic polymer LiClO$_4$-PEO, figure \ref{fig:GFET_TMDs}e \cite{Mehew2017}. In this structure, WS$_2$ is the light absorbing layer whilst the LiClO$_4$-PEO acts as a flexible and transparent top gate. Interestingly this device can be operated at bandwidths up to $1.5\,\mathrm{kHz}$ whilst maintaining a gain in excess of $10^6$ (figure \ref{fig:GFET_TMDs}f) without the need for gate pulsing or implementation of a photodiode structure. Highly mobile ions of the ionic polymer screen charge traps present in the device resulting in sub-millisecond response times and a responsivity of $10^6\,\mathrm{A/W}$ \cite{Mehew2017}. Lu \textit{et al} \cite{Lu2016} developed a vacuum annealing process to eliminate the trap states formed at the interface between graphene and GaSe nanosheets.

In place of a sensitising material, inserting a thin tunnel barrier between two graphene sheets can give rise to a large gain mechanism as photoexcited hot carriers generated in the top layer tunnel into the bottom layer. Liu \textit{et al} \cite{Liu2014} reported an ultra-broadband spectral response with responsivity $\sim1\,\mathrm{A/W}$ in the MIR increasing to $10^3\,\mathrm{A/W}$ in the visible using a $5$-nm-thick Ta$_2$O$_5$ layer sandwiched between two CVD-grown graphene monolayers. The operating bandwidth of this device is $35\,\mathrm{Hz}$ and with a NEP of $1\times10^{-11}\,\mathrm{WHz^{-1/2}}$ it is possible to estimate a LDR of $15\,\mathrm{dB}$.

\section{Summary and future outlook}
The development of ultra-thin, flexible and transparent photodetectors is ongoing. Graphene and other two-dimensional materials have enabled a new class of devices to be developed. This article presents an overview of the photodetector technologies based on chemically-functionalised graphene and hybrid structures such as graphene/QDs interfaces and graphene/TMDs heterostructures. 

Pristine graphene has a broadband absorption and fast response dominated by hot-carrier dynamics. The limited responsivity of these devices is related to the low absorption of single-layer graphene and to the fast relaxation time which does not allow carriers multiplication. To enhance the absorption and to improve the responsivity, chemical functionalisation has been used to modify the properties of pristine graphene. Although functionalised graphene PDs show a small improvement in responsivity and a drop in operating speed (bandwidth) compared to pristine graphene, chemical functionalisation allows photodetection from UV to THz wavelengths. Furthermore, quenching of PTE results in an increase of the LDR of such detectors. Chemical functionalisation also allows the creation of solution-processed materials, such as GO, with clear advantages in scalability.

Improvements in responsivity and operating bandwidth have been achieved by combining graphene with other materials to form hybrid photodetectors. Photodetectors based on hybrid interfaces of graphene with QDs, semiconductors to include atomically thin TMDs, perovskites and organic crystals offer improvements in responsivity and high gain owing to the photogating effect which enables charge multiplication in the graphene channel (charge recirculation). The majority of these devices have a limited LDR due to the charge relaxation time which quickly saturates the available states for photoexcitation, leading to a drop in responsivity with incident optical power. In some architectures, however, this effect is compensated by the low NEP, giving both high LDR and high responsivity.  Therefore, a thorough investigation of charge trapping mechanisms is necessary to design a high performance PD. The speed of these devices is limited by the charge trapping times with reported operating bandwidths between $\sim1\,\mathrm{Hz}$ and $\sim10\,\mathrm{kHz}$. Other limitations of biased graphene detectors are the high noise levels and power consumption given the large dark current present.

Future developments of atomically thin PDs will focus on the optimization of the responsivity and bandwidth for a given application. For instance, as quantum technologies become more important, the ability to manipulate single quanta of light is a priority and single-photon detectors will have to be integrated in optical communications systems and computation circuits at room temperature. Current state-of-the-art commercial single photon detectors have a dark count of $25\,\mathrm{electrons/s}$, which translates to a minimum detectable power of $0.14\,\mathrm{fW}$ (NEP) and a bandwidth exceeding $30\,\mathrm{MHz}$ \cite{Hadfield2009}. This corresponds to a detectivity of $1\times10^{17}\,\mathrm{Jones}$ assuming a $10\,\mu\mathrm{m}^2$ active area. Currently such performance has been approached in graphene-based detectors operating at low temperature ($T\sim100\,\mathrm{K}$) \cite{Roy2017}. Atomically-thin single-photon detectors, operating at room temperature, will allow a much faster integration into electronics and computation systems. Furthermore, thermal management of such devices will be facilitated by the very small footprint and low-energy consumption \cite{Song2018}.

Although high responsivity is an important parameter, in applications where high levels of illumination are present it is more important to have a large LDR, in order to avoid saturating a detector undergoing abrupt changes in radiation intensity. Such scenarios include surveillance or monitoring in harsh environments such as space or inside a nuclear reactor. For instance, recent efforts in nuclear fusion have been focussing in igniting a plasma using high-power lasers. In this case  monitoring the power of the laser and the corresponding plasma is vital and small-footprint photodetectors could significantly facilitate the operation of such reactors. FeCl$_3$-intercalated graphene detectors have shown a strong resilience when illuminated with high power densities (up to $100\,\mathrm{MW/cm^2}$) \cite{DeSanctis2017} and in harsh environments \cite{Wehenkel2015}. The large saturation power is due to the removal of bottlenecks in the cooling mechanism of hot carriers in graphene, enabled by the high levels of doping induced by the FeCl$_3$ molecules. This change in carrier dynamics increases the available density of states for photoexcitation and leads to a linear response with incident power \cite{DeSanctis2017}.

Healthcare applications will also benefit from ultra-thin, flexible photodetectors able to operate across a wide range of wavelengths \cite{DeSanctis20170057}. UV radiation, for example, is used for water purification and sterilisation. Control of the levels of illumination is important to enhance the efficiency of such techniques, especially if deployed in remote locations not served by an electricity grid. IR to THz radiation is used to perform spectroscopy measurements for chemical analysis. The development of portable and disposable spectrometers will allow to perform complex chemical analysis \textit{in-situ}, with evident benefits for fast and targeted response, for example, to an epidemic or environmental pollution. For these kind of applications, spectral selectivity and high responsivity are needed. Fast response time is not critical as the majority of analytical techniques have a time-scale much larger that the response time of the PDs.

In summary, graphene-based photodetectors, whilst offering a small footprint promising for next-generation flexible and wearable electronics, are also more energy-efficient and could provide features which are not available in bulk semiconductors, such as polarisation sensitivity \cite{Aslan2016} and strain-tunable response \cite{DeSanctis2018}. The family of more than 2000 layered materials offer a wealth of possibilities for the realisation of novel optoelectronic devices which can integrate multiple functionalities in a single active material.

\vspace{6pt} 

\authorcontributions{A.De-S. and J.D.M. collected and analysed the data, wrote the first draft and prepared the figures. All authors contributed to the final revision.}

\funding{M.F.C. and S.R. acknowledge financial support from: Engineering and Physical Sciences Research Council (EPSRC) of the United Kingdom, projects EP/M002438/1, EP/M001024/1, EPK017160/1, EP/K031538/1, EP/J000396/1; the Royal Society, grant title "Room temperature quantum technologies" and "Wearable graphene photovolotaic"; Newton fund, Uk-Brazil exchange grant title "Chronographene" and the Leverhulme Trust, research grants "Quantum drums" and "Quantum revolution". J.D.M. acknowledges financial support from the Engineering and Physical Sciences Research Council (EPSRC) of the United Kingdom, via the EPSRC Centre for Doctoral Training in Metamaterials, Grant No. EP/L015331/1.}

\conflictsofinterest{The authors declare no conflict of interest.} 

\abbreviations{The following abbreviations are used in this manuscript (in alphabetic order):\\

\noindent 
\begin{tabular}{@{}ll}
CNP	& Charge neutrality point\\
EQE	& External quantum efficiency\\
FET	& Field-effect transistor\\
FG	& Fluorographene\\
FLG	& Few-layer graphene\\
GO	& Graphene oxide\\
hBN	& Hexagonal boron nitride\\
IQE	& Internal quantum efficiency\\
IR	& infra-red\\
LDR	& Linear dynamic range\\
LED	& Light-emitting diode\\
MIR	& mid infra-red\\
NEP	& Noise Equivalent Power\\
NIR	& near infra-red\\
PD(s)	& Photodetector(s)\\
PG	& Photogating\\
PSD	& Posision-sensitive (photo)detector\\
PTE	& Photo-thermoelectric effect\\
PV	& Photovoltaic\\
QD(s)	& Quantum dot(s)\\
rGO	& reduced graphene oxide\\
SPCM	& Scanning-photocurrent map(ing)\\
TMD(s)	& Transition-metal dichalcogenide(s)\\
UV	& ultra-violet\\
vdW	& van der Waals\\

\end{tabular}}

\externalbibliography{yes}
\bibliography{Review_Materials_ADeS_Biblio}

\end{document}